\documentclass[11pt,nofootinbib,preprint,superscriptaddress]{revtex4}
\usepackage{epsfig,subfigure}
\usepackage{epstopdf}
\usepackage[utf8]{inputenc}
\usepackage{graphicx}
\usepackage{amssymb}
\usepackage{amsxtra}
\usepackage{amsmath}
\usepackage{booktabs,multirow,tabularx}
\usepackage{slashed}
\usepackage{float}
\usepackage{placeins}
\usepackage{rotating}
\usepackage{lscape}
\usepackage{color}
\usepackage{hyperref}
\usepackage{soul}
\usepackage{bm}

\newcommand{\bea}{\begin{eqnarray}}
\newcommand{\eea}{\end{eqnarray}}

\def\beq{\begin{equation}}
\def\eeq{\end{equation}}

\renewcommand{\Re}[1]{\textrm{Re}\left( #1 \right)}
\renewcommand{\Im}[1]{\textrm{Im}\left( #1 \right)}

\def\gev{\rm GeV}
\def\tev{\rm TeV}

\begin{document}

\title{Resonant Multi-Scalar Production in the Generic Complex Singlet Model in the Multi-TeV Region\\\vspace{-0.16in}}

\author{Samuel D. Lane}
\email{samuellane@upike.edu}
\affiliation{Division of Math and Natural Sciences, University of Pikeville, Pikeville, Kentucky, 41501~ U.S.A.}
\affiliation{Department of Physics, KAIST, Daejeon, 34141~ Korea}
\affiliation{Department of Physics and Astronomy, University of Kansas, Lawrence, Kansas, 66045~ U.S.A.}

\author{Ian M. Lewis}
\email{ian.lewis@ku.edu}
\affiliation{Department of Physics and Astronomy, University of Kansas, Lawrence, Kansas, 66045~ U.S.A.}

\author{Matthew Sullivan}
\email{msullivan1@bnl.gov}
\affiliation{High Energy Theory Group, Physics Department, Brookhaven National Laboratory,
Upton, New York, 11973~ U.S.A.}

\begin{abstract}
We develop benchmarks for resonant di-scalar production in the generic complex singlet scalar extension of the Standard Model (SM) with no additional symmetries, which contains two new scalars.  These benchmarks maximize di-scalar resonant production modes at future $pp$ colliders: $pp\rightarrow h_2 \rightarrow h_1 h_1$,  $pp\rightarrow h_2 \rightarrow h_1 h_3$, and $pp\rightarrow h_2 \rightarrow h_3 h_3$, where $h_1$ is the observed SM-like Higgs boson and $h_{2,3}$ are new scalars.  The decays $h_2\rightarrow h_1h_3$ and $h_2\rightarrow h_3h_3$ may be the only way to discover $h_3$, leading to a discovery of two new scalars at once.    Current LHC and projected future collider (HL-LHC, FCC-ee+HL-LHC, ILC500+HL-LHC) constraints on this model are used to produce benchmarks at the HL-LHC for $h_2$ masses between $250$ GeV and 1 TeV and a future $pp$ collider (FCC-hh) for $h_2$ masses between $250$~GeV and 12~TeV. We update the current LHC bounds on the singlet-Higgs boson mixing angle for these benchmarks.  As the mass of $h_2$ approaches the multi-TeV region, certain limiting behaviors of the maximum rates are uncovered due to theoretical constraints on the parameters. These limits, which can be derived analytically, are ${\rm BR}(h_2\rightarrow h_1h_1)\rightarrow 0.25$, ${\rm BR}(h_2\rightarrow h_3h_3)\rightarrow 0.5$, and ${\rm BR}(h_2\rightarrow h_1h_3) \rightarrow 0$. It can also be shown that the maximum rates of $pp\rightarrow h_2\rightarrow h_1h_1$ and $pp\rightarrow h_2\rightarrow  h_3h_3$ approach the same value.  Hence, all three $h_2\rightarrow h_ih_j$ decays are promising discovery modes for $h_2$ masses at and below $\mathcal{O}(1~{\rm TeV})$, while above $\mathcal{O}(1~{\rm TeV})$ the decays $h_2\rightarrow h_1h_1$ and $h_2\rightarrow h_3h_3$ are more encouraging. 
We choose benchmark masses for $h_3$ to produce a large range of decay signatures including multi-b, multi-vector boson, and multi-SM-like Higgs production. As we will show, the behavior of the maximum rates leads to the surprising conclusion that in the multi-TeV region this model may be discovered in the Higgs quartet production mode via $h_2\rightarrow h_3h_3\rightarrow 4\,h_1$ decays before Higgs triple production is observed.  The maximum di- and four Higgs production rates are similar in the multi-TeV range.  
\end{abstract}
\maketitle

\section{Introduction}
Measuring and understanding the properties of the Higgs boson is one the long term goals of the LHC and future collider programs~\cite{Dawson:2022zbb}.  One of the biggest questions is the exact form of the scalar potential and the nature of electroweak symmetry breaking (EWSB).  In the Standard Model (SM), there are only two parameters in the Higgs potential.  Hence, once the Higgs vacuum expectation value (vev) is set, $v_{\rm EW}=246$~GeV, and the Higgs mass is known, $m_h=125$~GeV~\cite{ATLAS:2022vkf,CMS:2022dwd}, the potential parameters can be solved for and the SM is completely predictive.  Since EWSB originates from the Higgs potential, to determine that EWSB is SM-like the shape of the scalar potential must be probed.  So far, only the quadratic mass term of the Higgs potential has been measured directly.  The self-couplings of the Higgs must be determined to more fully explore the shape of the scalar potential.  At hadron colliders, di-Higgs production~\cite{Djouadi:1999rca,Glover:1987nx,Plehn:1996wb,Plehn:2001nj,DiMicco:2019ngk,Dawson:1998py}, which depends on the Higgs tri-linear coupling, is the most direct method to measure the Higgs self-couplings and test the SM predictions about the nature of EWSB.

In the SM, the di-Higgs production from gluon fusion at the $\sqrt{S}=13$~TeV LHC is predicted to be $\sigma=31.05$~fb~\cite{Dawson:2022zbb,Dawson:1998py,Borowka:2016ehy,Baglio:2018lrj,Grazzini:2018bsd,Baglio:2020wgt}.  Due to its small rate, double Higgs production can be quite sensitive to new physics contributions~\cite{DiMicco:2019ngk}. 
Of particular interest are scalar extensions of the SM that can change the nature of EWSB and the Higgs self-couplings.  Indeed, there has been interest investigating scenarios in which the first evidence of new physics could be through variations of the Higgs tri-linear couplings~\cite{DiLuzio:2017tfn,Chang:2019vez,Bahl:2022jnx,Durieux:2022hbu}.  Another spectacular signature of additional scalars could be resonant di-Higgs production.  Such a signal is much studied and appears in even the simplest of all possible extensions: a real scalar that is a singlet under the SM gauge group~\cite{Silveira:1985rk,OConnell:2006rsp,Barger:2007im,Bowen:2007ia,Profumo:2007wc,
No:2013wsa,Pruna:2013bma,Profumo:2014opa,Chen:2014ask,Dawson:2015haa,Buttazzo:2015bka,Robens:2015gla,
Robens:2016xkb,Kotwal:2016tex,Lewis:2017dme,Huang:2017jws,Dawson:2018dcd,Li:2019tfd,Alves:2020bpi,Papaefstathiou:2020iag,Dawson:2020oco,Adhikari:2020vqo,Papaefstathiou:2021glr,Dawson:2021jcl,Hammad:2023sbd}.

There has been recent interest in moving beyond di-Higgs production and studying more generic di-scalar production.  For example, non-resonant production of two new scalars can provide insight into the early universe electroweak phase transition~\cite{Curtin:2014jma,Chen:2017qcz}.  A more spectacular signal is asymmetric resonant production of the SM Higgs+a new scalar or symmetric resonant production of two new scalars~\cite{Basler:2018dac,Abouabid:2021yvw,Adhikari:2022yaa,Dawson:2017jja,Costa:2015llh,Robens:2019kga,Papaefstathiou:2020lyp}.  In this paper, we develop benchmark points for such signals in the SM extended by a complex scalar that is a singlet under the SM gauge group~\cite{Coimbra:2013qq,Costa:2014qga,Ferreira:2016tcu,Muhlleitner:2017dkd,Basler:2019nas,Muhlleitner:2020wwk,Egle:2022wmq,Egle:2023pbm,Barger:2008jx,Alexander-Nunneley:2010tyr,Dawson:2017jja,Adhikari:2022yaa,Costa:2015llh,Robens:2019kga,Papaefstathiou:2020lyp,Dawson:2023oce}. Beyond providing interesting collider signatures of new scalars, the complex singlet extension can also help provide a strong first order electroweak phase transition necessary to generate the baryon asymmetry of the universe via electroweak baryogenesis~\cite{Barger:2008jx,Chao:2014ina,Jiang:2015cwa,Chiang:2017nmu,Cheng:2018ajh,Grzadkowski:2018nbc,Chen:2019ebq,Chiang:2019oms,Cho:2021itv,Cho:2022our,Cho:2022zfg,Zhang:2023mnu}. This extension has been much studied in the scenarios where there are additional $\mathbb{Z}_2$ or $U(1)$ symmetries~\cite{McDonald:1993ex,Cerdeno:2006ha,Barger:2008jx,Coimbra:2013qq,Costa:2014qga,Costa:2015llh,Ferreira:2016tcu,Muhlleitner:2017dkd,Basler:2019nas,Muhlleitner:2020wwk,Robens:2019kga,Papaefstathiou:2020lyp,Egle:2022wmq,Egle:2023pbm}.  We will focus on the scenario in which there is no additional symmetry beyond the SM gauge group~\cite{Dawson:2017jja,Adhikari:2022yaa}.

The complex singlet scalar extension with no additional symmetries is equivalent to the SM extended by two real singlet scalars~\cite{Dawson:2017jja}. 
Hence, it is in some sense a next-to-minimal extension of the SM in terms of field content and symmetry.  There are then three real scalars: $h_1,\,h_2,$ and $h_3$, where $h_1$ is identified with observed $125$~GeV scalar.  Hence, when kinematically allowed, there could be resonant production of multi-scalar final states:
\begin{eqnarray}
h_2\rightarrow h_1h_1,\quad h_2\rightarrow h_1h_3,\quad{\rm and}\quad h_2\rightarrow h_3h_3.
\end{eqnarray}
If the mixing between $h_3$ and the SM Higgs boson is minuscule, the coupling between $h_3$ and SM fermions and gauge bosons will be negligible. Hence, the production rate of $h_3$ through gluon fusion or vector boson fusion, $\sigma(pp \rightarrow h_3) = \sin^2 \theta_1 \sin^2\theta_2\sigma_{\rm SM}(pp\rightarrow h_3)$, is very small.  Indeed, for mixing angles $\sin \theta_1\leq 0.2$ and $\sin \theta_2 \leq 0.01$, which, as we will see later, are reasonable ranges, then  $\sigma(pp\rightarrow h_3) \leq 4 \times 10^{-6} \cdot \sigma_{\rm SM}(pp \rightarrow h_3 )$.  This is more than an order of magnitude below the expected 95\% CL direct search limits of the FCC-hh\cite{EuropeanStrategyforParticlePhysicsPreparatoryGroup:2019qin}. In that case, the major production mode of $h_3$ would then be directly through these decays.   Hence, it is possible that searches for resonant production of $h_2\rightarrow h_1h_3$ and $h_2\rightarrow h_3h_3$ could provide a discovery of two new real scalars at once, and could be the only feasible mechanisms to discover $h_3$~\cite{Dawson:2017jja}. 

Our benchmark points are determined by individually maximizing the resonant production rates of $pp\rightarrow h_2\rightarrow h_1h_1$, $pp\rightarrow h_2\rightarrow h_1h_3$, and $pp\rightarrow h_2\rightarrow h_3h_3$.  The benchmarks are constructed to be relevant for the HL-LHC as well as future higher energy hadron colliders.  As such, we will consider several scenarios for current and projected experimental constraints:
\begin{itemize}
\item S1: Current constraints from the LHC.
\item S2: Projected constraints at the high luminosity LHC (HL-LHC)~\cite{EuropeanStrategyforParticlePhysicsPreparatoryGroup:2019qin,Dawson:2022zbb}.
\item S3: Projected constraints at the HL-LHC and a future circular electron positron collider such as the FCC-ee or CEPC~\cite{EuropeanStrategyforParticlePhysicsPreparatoryGroup:2019qin,Dawson:2022zbb,FCC:2018evy,CEPCStudyGroup:2018rmc,CEPCStudyGroup:2018ghi}.
\item S4: Projected constraints at the HL-LHC and a 500 GeV International Linear Collider (ILC500)~\cite{EuropeanStrategyforParticlePhysicsPreparatoryGroup:2019qin,Dawson:2022zbb,Behnke:2013xla,ILC:2013jhg,Adolphsen:2013jya}.
\end{itemize}
Since scenario S1 concerns a 13-14~TeV LHC, we consider $h_2$ masses up to 1 TeV.  For the future collider projects, scenarios S2-S4, we will consider $h_2$ masses up to 12 TeV so that they are relevant even for a 100 TeV $pp$ collider.

In addition to the experimental constraints, we will consider theory constraints such as the scalar potential having the correct EWSB pattern to create the known particle masses, the potential is bounded from below, perturbative unitarity, and that the total width of the heavy resonance is narrow, i.e. less than 10\% of the resonance mass.  As we will show analytically and numerically, in the multi-TeV regime these theory constraints have strong implications.  These constraints cause the maximum rates of $pp\rightarrow h_2\rightarrow h_1h_1$ and $pp\rightarrow h_2\rightarrow h_3h_3$ to converge to the same value at large $h_2$ masses, even though the branching ratios converge to different values: ${\rm BR}(h_2\rightarrow h_1 h_1)\approx 0.25$ and ${\rm BR}(h_2\rightarrow h_3h_3)\approx 0.5$.  Additionally, for $h_2\rightarrow h_1h_3$, the theory constraints, in particular having the correct global minimum for the scalar potential and perturbative unitarity, result in ${\rm BR}(h_2\rightarrow h_1h_3)\rightarrow 0$ as the mass of $h_2$ increases. We derive an analytical understanding of these results.

The structure of the paper is as follows.  In section \ref{sec:model} we introduce the scalar potential and discuss the model in detail. In section \ref{sec:Const} we cover the various theoretical and experimental constraints on the model, including updated bounds on the scalar mixing angle. We present our results and a discussion of the collider phenomenology of multi-Higgs signals in section \ref{sec:benchmark}.  Finally, we conclude in section~\ref{sec:conc}.  Details of the perturbative unitarity constraints are in given in appendix~\ref{app:PertUnit}, theoretical bounds on cubic and quartic potential parameters are derived in appendix~\ref{app:CubicBounds}  as well as trilinear couplings in the mass basis and $h_2$ branching ratios in appendix~\ref{app:TrilinearBounds}, and experimental constraints and searches are discussed in detail in appendix~\ref{app:lims}.

\section{Model}
\label{sec:model}
The model under consideration consists of the SM extended by a gauge singlet complex scalar, $S_c$.  Many previous studies of the complex singlet extension have imposed a softly broken $\mathbb{Z}_2$ or $U(1)$ symmetry on the new scalar singlet~\cite{Coimbra:2013qq,Costa:2014qga,Ferreira:2016tcu,Muhlleitner:2017dkd,Basler:2019nas,Muhlleitner:2020wwk,Egle:2022wmq,Egle:2023pbm,Barger:2008jx,Alexander-Nunneley:2010tyr,Dawson:2017jja,Adhikari:2022yaa,Costa:2015llh,Robens:2019kga,Papaefstathiou:2020lyp,Dawson:2023oce}. We follow Ref.~\cite{Dawson:2017jja} and use the most general potential with no additional symmetry beyond the SM:
\begin{eqnarray}
V(\Phi,S_c)&=&-\frac{\mu^2}{2}\Phi^\dagger \Phi+\frac{\lambda}{4}(\Phi^\dagger\Phi)^2+\frac{b_2}{2}|S_c|^2+\frac{d_2}{4}|S_c|^4+\frac{\delta_2}{2}\Phi^\dagger\Phi |S_c|^2\nonumber\\
&&+\left(a_1\,S_c+\frac{b_1}{4}\,S_c^2+\frac{e_1}{6}\,S_c^3+\frac{e_2}{6}S_c|S_c|^2+\frac{\delta_1}{4} \Phi^\dagger\Phi\,S_c+\frac{\delta_3}{4}\Phi^\dagger\Phi\,S_c^2\right.\nonumber\\
&&\left.+\frac{d_1}{8}S_c^4+\frac{d_3}{8}S_c^2|S_c|^2+{\rm h.c.}\right)~\label{eq:VSc}
\end{eqnarray}
where $S_c=(S_0+i\,A)/\sqrt{2}$ is the complex singlet scalar, $\mu^2,\lambda,b_2,d_2,$ and $\delta_2$ are real parameters, and $a_1,b_1,e_1,e_2,\delta_1,\delta_3,d_1,$ and $d_3$ are complex parameters.  In the most general potential, the vev of the complex singlet can be set to zero, $\langle S_c\rangle =0$, without loss of generality~\cite{Dawson:2017jja,Chen:2014ask}.  This can be understood by noting that the vev of $S_c$ breaks no symmetry and a shift $S_c\rightarrow S_c+\langle S_c\rangle$ introduces no new interactions.  That is, the shift is a non-physical redefinition of parameters and we have the freedom to set $\langle S_c\rangle=0$. The Higgs doublet is denoted as $\Phi$ and it takes the form
\begin{eqnarray}
\Phi=\frac{1}{\sqrt{2}}\begin{pmatrix} \sqrt{2}G^+\\v_{\rm EW}+h+i\,G^0\end{pmatrix},
\end{eqnarray}
where $v_{\rm EW}=246$~GeV is the Higgs vev, $h$ is the SM Higgs boson, $G^+$ is the charged Goldstone boson, and $G^0$ is the neutral Goldstone boson.

The complex singlet introduces two new CP even scalars: $S_0$ and $A$.  The three neutral scalars $h,\,S_0,$ and $A$ then mix into three scalar mass eigenstates via an $SO(3)$ rotation:
\begin{eqnarray}
\begin{pmatrix} h_1\\h_2\\h_3\end{pmatrix} =\begin{pmatrix} \cos\theta_1 & -\sin\theta_1 & 0\\\sin\theta_1\cos\theta_2 & \cos\theta_1\cos\theta_2 & \sin\theta_2 \\ \sin\theta_1\sin\theta_2 & \cos\theta_1 \sin\theta_2 & - \cos\theta_2\end{pmatrix}\begin{pmatrix} h\\S_0\\A\end{pmatrix},\label{eq:mixing}
\end{eqnarray}
where the mass eigenstates are $h_1,\,h_2,\,$ and $h_3$ with masses $m_1,\,m_2,\,$ and $m_3$, respectively.  We will associate $h_1$ with the observed scalar and set $m_1=125$~GeV~\cite{ATLAS:2022vkf,CMS:2022dwd}.  There is, in principle, a third $SO(3)$ rotation angle in Eq.~(\ref{eq:mixing}), but it can be removed via a phase rotation of $S_c\rightarrow e^{i\,\theta_3}S_c$~\cite{Dawson:2017jja}.  Such a phase rotation is at most a redefinition of parameters in the scalar potential [Eq.~(\ref{eq:VSc})].  Hence, the third rotation angle can be removed with no physical effect. The free parameters of the model are then
\begin{eqnarray}
v=246~{\rm GeV},\,\langle S_c\rangle =0,m_1=125~{\rm GeV},\,m_2,\,m_3,\,\theta_1,\,\theta_2,\,\theta_3=0,\,\delta_2,\,\delta_3,\,d_1,\,d_2,\,d_3,\,e_1,\,e_2.
\end{eqnarray}
By requiring that $(\langle \Phi^\dagger\Phi\rangle,\langle S_c\rangle)=(v_{\rm EW}^2/2,0)$ be an extremum of the potential and going to the scalar mass basis, the Lagrangian parameters can be related to these free parameters according to the following equations:
\begin{eqnarray}
    \mu^2 &=& m_1^2 \cos^2{\theta_1} + (m_2^2 \cos^2{\theta_2} + m_3^2 \sin^2{\theta_2})\sin^2{\theta_1}, \\
    \lambda &=& \frac{2(m_1^2 \cos^2{\theta_1} + (m_2^2 \cos^2{\theta_2} + m_3^2 \sin^2{\theta_2})\sin^2{\theta_1})}{v_{\rm EW}^2}, \\
    a_1 &=& \frac{(m_1^2 - m_2^2 \cos^2{\theta_2} - m_3^2 \sin^2{\theta_2}) v_{\rm EW} \sin{2\theta_1}}{4 \sqrt{2}} + i \frac{(m_2^2 - m_3^2) v_{\rm EW} \sin{\theta_1} \sin{2 \theta_2}}{4 \sqrt{2}}, \\
    \delta_1 &=& -\frac{\sqrt{2} (m_1^2 - m_2^2 \cos^2{\theta_2} - m_3^2 \sin^2{\theta_2}) \sin{2 \theta_1}}{v_{\rm EW}} - i \frac{\sqrt{2} (m_2^2 - m_3^2) \sin{\theta_1} \sin{2 \theta_2}}{v_{\rm EW}}, \\
    b_1 &=& -\frac{1}{2} v_{\rm EW}^2 \delta_3 + m_1^2 \sin^2{\theta_1} + m_2^2 (\cos^2{\theta_1} \cos^2{\theta_2} - \sin^2{\theta_2}) + m_3 (\cos^2{\theta_1}\sin^2{\theta_2}-\cos^2{\theta_2}) \nonumber \\
    &+&  i (m_3^2 - m_2^2) \cos{\theta_1}\sin{2\theta_2}, \\
    b_2 &=& -\frac{v_{\rm EW}^2 \delta_2}{2} + m_1^2 \sin^2{\theta_1} + m_2^2 (\cos^2{\theta_1} \cos^2{\theta_2} + \sin^2{\theta_2}) + m_3^2 (\cos^2{\theta_1} \sin^2{\theta_2} + \cos^2{\theta_2}).
\end{eqnarray}

We will be interested in the limit $|\theta_2|\ll 1$.  The mixing matrix of Eq.~(\ref{eq:mixing}) then becomes
\begin{eqnarray}
\begin{pmatrix} h_1\\h_2\\h_3\end{pmatrix} =\begin{pmatrix} \cos\theta_1 & -\sin\theta_1 & 0\\\sin\theta_1 & \cos\theta_1 & \sin\theta_2 \\ \sin\theta_1\sin\theta_2 & \cos\theta_1 \sin\theta_2 & - 1\end{pmatrix}\begin{pmatrix} h\\S_0\\A\end{pmatrix}+\mathcal{O}(\sin^2\theta_2)\label{eq:mixinglim}
\end{eqnarray}
In this limit, $h_2$ couples to fermions and gauge bosons like a SM Higgs boson of mass $m_2$ but suppressed by $\sin\theta_1$, independent of whether $\theta_2$ or $\theta_1$ is larger.  
The couplings of $h_3$ to fermions and gauge boson are doubly suppressed by $\sin\theta_1\sin\theta_2$.  As a result, when kinematically allowed, we expect $h_3$'s main production mechanism to be via decays of $h_2$: $h_2\rightarrow h_1h_3$ or $h_2\rightarrow h_3h_3$.  With these considerations, for our benchmark points we will consider the mass ordering $m_2>m_3>m_1$.

The trilinears relevant for the channels we will consider show up in the potential as
\begin{eqnarray}
\label{eq:trilinearpotential}
V\supset \frac{1}{2}\lambda_{112}h_1^2h_2+\lambda_{123}h_1h_2h_3+\frac{1}{2}\lambda_{233}h_2h_3^2 + \frac{1}{2} \lambda_{113} h_1^2 h_3.
\end{eqnarray}
The partial widths of $h_2$ to $h_1 h_1$, $h_1 h_3$, and $h_3 h_3$ are given in terms of the masses and trilinears by
\begin{eqnarray}
\label{eq:h1h1width}
\Gamma(h_2\rightarrow h_1h_1) &=& \frac{\lambda^2_{112} }{32 \pi m_2}\sqrt{1 -  \frac{4\,m_1^2}{m_2^2}}, \\
\label{eq:h1h3width}
\Gamma(h_2\rightarrow h_1h_3) &=& \frac{\lambda^2_{123} }{16 \pi m_2} \sqrt{\left[1-\left(\frac{m_1-m_3}{m_2}\right)^2\right]\left[1-\left(\frac{m_1+m_3}{m_2}\right)^2\right]},\\
\label{eq:h3h3width}
\Gamma(h_2\rightarrow h_3h_3) &=& \frac{\lambda^2_{233} }{32 \pi m_2}\sqrt{1-\frac{4\,m_3^2}{m_2^2}}.
\end{eqnarray}
Note that the coefficient of Eq.~(\ref{eq:h1h3width}) differs by a factor of 2 from the other partial widths due to the two final state scalars not being identical. The general expressions for the trilinears in Eq.~(\ref{eq:trilinearpotential}) are complicated, but using Eqs.~(\ref{eq:VSc},\ref{eq:mixinglim}), the $\theta_2\rightarrow 0$ limits of the relevant trilinear couplings are
\begin{eqnarray}
\label{eq:trilinear112}
\lambda_{112}&=&\sin\theta_1\frac{m_2^2}{v_{\rm EW}}\left[\cos^2\theta_1\left(1+2\frac{m_1^2}{m_2^2}\right)+\frac{1}{\sqrt{2}}\cos\theta_1\sin\theta_1\left(\Re{e_1}+\Re{e_2}\right)\frac{v_{\rm EW}}{m_2^2}\right.\\
&&\left.-\left(1-\frac{3}{2}\sin^2\theta_1\right)\frac{v_{\rm EW}^2}{m_2^2}\left(\delta_2+\Re{\delta_3}\right)\right]\nonumber\\
\label{eq:trilinear123}
\lambda_{123}&=&-\frac{\sqrt{2}}{6}\cos\theta_1\sin\theta_1\left(3\,\Im{e_1}+\Im{e_2}\right)+\frac{1}{2}\left(\cos^2\theta_1-\sin^2\theta_1\right)\Im{\delta_3}\,v_{\rm EW}\\
\label{eq:trilinear233}
\lambda_{233}&=&\frac{\sqrt{2}}{6}\cos\theta_1\left(\Re{e_2}-3\,\Re{e_1}\right)+\frac{1}{2}\sin\theta_1\,v_{\rm EW}\left(\delta_2-\Re{\delta_3}\right).
\end{eqnarray}
When kinematically accessible, the partial width of $h_3$ to $h_1 h_1$ is given similarly by 
\begin{equation}
\label{eq:h3toh1h1width}
\Gamma(h_3\rightarrow h_1h_1) = \frac{\lambda^2_{113} }{32 \pi m_3}\sqrt{1 -  \frac{4\,m_1^2}{m_3^2}} ,
\end{equation}
with the trilinear coupling, expanding in small $\theta_2$, being 
\begin{eqnarray}
\lambda_{113} &=& \frac{1}{6} \sin{\theta_1} \left[\sqrt{2} \left(3 \Im{e_1}+\Im{e_2}\right) \sin{\theta_1}-6 v_{\rm EW} \cos{\theta_1} \Im{\delta_3}\right] \nonumber \\
   &+& \frac{1}{4} \Bigg[ \Bigg. 4 \cos^2{\theta_1} \frac{2m_1^2+m_3^2}{v_{\rm EW}} + \sqrt{2} \left(\Re{e_1}+\Re{e_2}\right)\sin{2\theta_1} \nonumber \\
   &-&  v_{\rm EW}\left(1+3\cos{2\theta_1}\right)\left(\Re{\delta_3}+\delta_2\right) \Bigg. \Bigg]\sin\theta_1\sin \theta_2+\mathcal{O}(\sin^2\theta_2)\label{eq:trilinear113} .
\end{eqnarray}

\section{Constraints}
\label{sec:Const}
Now we give the various constraints on the parameter space of this model.  First, we will cover an overview of the various theoretical constraints and then the current experimental constraints from precision Higgs measurements and direct searches for heavy new scalars.
\subsection{Theoretical Constraints}
\subsubsection{Boundedness and Global Minima}

The scalar potential must be bounded from below to stabilize the scalar fields against runaway directions.   That is, as the scalar fields approach infinity the potential must be positive.  At large field values, the quartic couplings dominate.  Hence, boundedness requires
\begin{eqnarray}
\frac{\lambda}{4}(\Phi^\dagger\Phi)^2+\frac{d_2}{2}|S_c|^4+\frac{\delta_2}{2}\Phi^\dagger\Phi |S_c|^2+\left(\frac{\delta_3}{4}\Phi^\dagger\Phi S_c^2+\frac{d_1}{8}S^4_c+\frac{d_3}{8}S_c^2|S_c|^2+{\rm h.c.}\right)\geq 0.
\end{eqnarray}
This condition is checked numerically for all directions in field space.  However, as discussed in App.~\ref{app:CubicBounds}, there are some directions that can give useful analytical bounds on the quartic coupling.  We note that although boundedness is checked numerically, since the generic complex scalar extension is equivalent to two real scalars, the scalar potential has complicated but known analytic conditions for vacuum stability~\cite{Kannike:2016fmd}.

With the additional terms added to the scalar potential as compared to the SM case, there are many potential extrema.  In Sec.~\ref{sec:model}, potential parameters are chosen such that one of the minima lies at 
\begin{eqnarray}
\langle \Phi\rangle = \begin{pmatrix} 0\\v_{\rm EW}/\sqrt{2}\end{pmatrix},\quad \langle S_c\rangle = 0.\label{eq:EWSBmin}
\end{eqnarray}
Since the singlet does not contribute to $W/Z$ or fermion masses, the other minima cannot reproduce the measured masses of SM particles.  Hence, to have the correct EWSB pattern, the minimum in Eq.~(\ref{eq:EWSBmin}) is required to be the global minimum.  As with bounded from below, this condition is checked numerically and at tree level.  It is possible to find some necessary but not sufficient analytic conditions to satisfy the global minimum constraint.  These conditions can place important bounds on quartic and cubic scalar couplings, as we review in Apps.~\ref{app:CubicBounds} and~\ref{app:TrilinearBounds}.

\subsubsection{Perturbative Unitarity}

Enforcing perturbative unitarity~\cite{Lee:1977eg, Lee:1977yc, Chanowitz:1978uj, Chanowitz:1978mv} on our parameter space helps guarantee that the parameters we consider are perturbative~\cite{Schuessler:2007av}. Such a requirement helps stabilize our conclusions against higher order corrections.  We examine two-to-two scalar scattering processes in the high energy limit in order to derive bounds on the quartic couplings. The partial wave expansion of each matrix element is
\begin{equation}
\mathcal{M}=16\pi\sum_{j=0}^{\infty} (2j+1)a_j P_j(\cos\theta)\label{eq:partialwave},
\end{equation}
where $P_j(\cos\theta)$ are the Legendre polynomials. The leading contributions in the high energy limit will be from the zero angular momentum $a_0$ term. We will treat both $\mathcal{M}$ and $a_0$ as matrices connecting different two-to-two scattering states.  The scattering matrix between electrically neutral initial and final states is then 
\begin{align}
&\mathcal{M} =\label{eq:matrix}\\
&\resizebox{\textwidth}{!}{$\begin{pmatrix}
  \frac{3 \lambda }{4} & 0 & 0 & \frac{\delta_2}{4}+\frac{\Re{\delta_3}}{4} &
   -\frac{\Im{\delta_3}}{2 \sqrt{2}} & \frac{\delta_2}{4}-\frac{\Re{\delta
  _3}}{4} & \frac{\lambda }{4} & \frac{\lambda }{2 \sqrt{2}} \\
 0 & \frac{\delta_2}{2}+\frac{\Re{\delta_3}}{2} & -\frac{\Im{\delta_3}}{2} & 0 & 0 & 0 & 0 & 0 \\
 0 & -\frac{\Im{\delta_3}}{2} & \frac{\delta_2}{2}-\frac{\Re{\delta
  _3}}{2} & 0 & 0 & 0 & 0 & 0 \\
 \frac{\delta_2}{4}+\frac{\Re{\delta_3}}{4} & 0 & 0 & \frac{3
   \Re{d_1}}{4}+\frac{3 \Re{d_3}}{4}+\frac{3 d_2}{4} & -\frac{3
   \Im{d_1}}{2 \sqrt{2}}-\frac{3 \Im{d_3}}{4 \sqrt{2}} &
   \frac{d_2}{4}-\frac{3 \Re{d_1}}{4} & \frac{\delta_2}{4}+\frac{\Re{\delta
  _3}}{4} & \frac{\delta_2}{2 \sqrt{2}}+\frac{\Re{\delta_3}}{2 \sqrt{2}} \\
 -\frac{\Im{\delta_3}}{2 \sqrt{2}} & 0 & 0 & -\frac{3 \Im{d_1}}{2
   \sqrt{2}}-\frac{3 \Im{d_3}}{4 \sqrt{2}} & \frac{d_2}{2}-\frac{3
   \Re{d_1}}{2} & \frac{3 \Im{d_1}}{2 \sqrt{2}}-\frac{3
   \Im{d_3}}{4 \sqrt{2}} & -\frac{\Im{\delta_3}}{2 \sqrt{2}} &
   -\frac{\Im{\delta_3}}{2} \\
 \frac{\delta_2}{4}-\frac{\Re{\delta_3}}{4} & 0 & 0 & \frac{d_2}{4}-\frac{3
   \Re{d_1}}{4} & \frac{3 \Im{d_1}}{2 \sqrt{2}}-\frac{3
   \Im{d_3}}{4 \sqrt{2}} & \frac{3 \Re{d_1}}{4}-\frac{3
   \Re{d_3}}{4}+\frac{3 d_2}{4} & \frac{\delta_2}{4}-\frac{\Re{\delta
  _3}}{4} & \frac{\delta_2}{2 \sqrt{2}}-\frac{\Re{\delta_3}}{2 \sqrt{2}} \\
 \frac{\lambda }{4} & 0 & 0 & \frac{\delta_2}{4}+\frac{\Re{\delta_3}}{4} &
   -\frac{\Im{\delta_3}}{2 \sqrt{2}} & \frac{\delta_2}{4}-\frac{\Re{\delta
  _3}}{4} & \frac{3 \lambda }{4} & \frac{\lambda }{2 \sqrt{2}} \\
 \frac{\lambda }{2 \sqrt{2}} & 0 & 0 & \frac{\delta_2}{2 \sqrt{2}}+\frac{\Re{\delta
  _3}}{2 \sqrt{2}} & -\frac{\Im{\delta_3}}{2}  & \frac{\delta_2}{2
   \sqrt{2}}-\frac{\Re{\delta_3}}{2 \sqrt{2}} & \frac{\lambda }{2 \sqrt{2}} &
   \lambda  \\
\end{pmatrix}$}\nonumber
\end{align} 
with the (normalized~\cite{Lee:1977eg, Lee:1977yc}) two-particle states being, in order, $\frac{hh}{\sqrt{2}}$, $hS$, $hA$, $\frac{SS}{\sqrt{2}}$, $AS$, $\frac{AA}{\sqrt{2}}$, $\frac{G^0G^0}{\sqrt{2}}$, and $G^+ G^-$. There is also a scattering matrix between electrically charged initial and final states.  However, it does not introduce any new constraints in addition to the scattering from neutral initial states to neutral final states.

The standard perturbative unitarity bound at tree-level is for the magnitude of the eigenvalues of $a_0$ to be less than $\frac{1}{2}$.  This is equivalent to a requirement that the minimum higher order corrections to these scattering processes be less than or equal to $41\%$~\cite{Schuessler:2007av}.  The upper bound on the magnitude of the eigenvalues of $a_0$ gives an upper bound of $8\pi$ on the eigenvalues of the  matrix $\mathcal{M}$ in Eq.~(\ref{eq:matrix}). By considering submatrices of $\mathcal{M}$, we obtain conservative, necessary but not sufficient bounds, on the quartic couplings:
\begin{eqnarray}
\left| \lambda \right| &\leq& \frac{16 \pi}{3}, \nonumber \\
\left|\delta_2\right|,\left|\Re{\delta_3}\right|,\left|\Im{\delta_3}\right| &\leq& 8 \sqrt{2} \pi,  \nonumber \\
\left|\Re{d_1}\right|, \left|\Im{d_1}\right| &\leq& \frac{16 \pi}{3}, \nonumber \\
\left| d_2 \right| &\leq& 8 \pi, \nonumber \\
\left|\Re{d_3}\right|, \left|\Im{d_3}\right| &\leq& \frac{32 \pi}{3}.
\label{eq:quarticbounds}
\end{eqnarray}
Notably, the bound on $\lambda$ in Eq.~(\ref{eq:quarticbounds}) places a constraint on the allowed masses and mixings. For a particular choice of mixing angles, this induces an upper limit on the allowed heavy Higgs masses.  In practice, as with the boundedness and global minimum constraints, we check numerically that the eigenvalues of Eq.~(\ref{eq:matrix}) are bounded by $8\pi$.

\subsection{Experimental Constraints}
\label{sec:exconst}

We now consider experimental constraints on this model.  First, we take Higgs precision measurements into account \cite{ATLAS:2022vkf,CMS:2022dwd,ATLAS-CONF-2022-067,CMS-PAS-HIG-19-011}.  These measurements are typically given in terms of signal strengths, i.e. ratios of beyond SM (BSM) and SM predictions for different Higgs production and decay channels:
\begin{eqnarray}
\mu_i^f=\frac{\sigma_i(pp\rightarrow h_1)}{\sigma_{i,{\rm SM}}(pp\rightarrow h_1)} \frac{{\rm BR}(h_1\rightarrow f)}{{\rm BR}_{\rm SM}(h_1\rightarrow f)},
\end{eqnarray}
where the subscript ${\rm SM}$ indicates SM predictions, quantities without the ${\rm SM}$ subscript are BSM predictions, $i$ is the initial state, and $f$ is the final state.  As can be seen from Eq.~(\ref{eq:mixing}), the $h_1$ couplings to fermions and gauge bosons are the same as the SM but suppressed by a universal factor of $\cos\theta_1$.  Hence, all $h_1$ rates are suppressed by $\cos^2\theta_1$.  As a consequence, the $h_1$ branching ratios are unchanged from the SM predictions and production cross sections scale as $\cos^2\theta_1$:
\begin{eqnarray}
{\rm BR}(h_1\rightarrow f)={\rm BR}_{\rm SM}(h_1\rightarrow f),\quad\sigma_i(pp\rightarrow h_1)=\cos^2\theta_1\, \sigma_{i,SM}(pp\rightarrow h_1).
\end{eqnarray}
  The signal strengths then become
\begin{eqnarray}
\mu_i^f=\cos^2\theta_1.\label{eq:sigstrength}
\end{eqnarray}
Note that these conclusions are independent of the choice of the other mixing angle $\theta_2$.  The mixing angle $\theta_1$ can then be fit using a $\chi^2$ distribution:
\begin{eqnarray}
\chi^2_{h_1}=\sum_{i,f}\frac{(\mu_i^f-\hat{\mu}_i^f)^2}{(\delta_i^f)^2},\label{eq:Chi2h1}
\end{eqnarray}
where $\hat{\mu}_i^f$ is the measured signal strength and $\delta_i^f$ is the uncertainty in the measured signal strengths (including systematic, statistical, and theoretical uncertainties).  

The signal strength in Eq.~(\ref{eq:sigstrength}) is the same for all initial and final states.  Hence, the global signal strengths that combine all channels~\cite{ATLAS:2022vkf,CMS:2022dwd} can be used. In that way, correlations between different initial and final states are accounted for.  However, since the global signal strengths were reported~\cite{ATLAS:2022vkf,CMS:2022dwd}, there have been updates to $h\rightarrow W^\pm W^{\mp,*}$ in the $Wh$ and $Zh$ production modes with $139$ fb$^{-1}$ at lab frame $pp$ energy $\sqrt{S}=13$~TeV from ATLAS~\cite{ATLAS-CONF-2022-067}.  CMS~\cite{CMS-PAS-HIG-19-011} updated $h\rightarrow b\bar{b}$ in the $t\bar{t}h$ production mode with $138$ fb$^{-1}$ at $\sqrt{S}=13$~TeV.
  Using the correlation matrices reported with the Higgs fit combinations~\cite{ATLAS:2022vkf,CMS:2022dwd}, we update the global signal strengths to
\begin{eqnarray}
\mu_{\rm ATLAS}=1.04\pm 0.06\quad{\rm and}\quad\mu_{\rm CMS}=0.96\pm0.06\label{eq:global}
\end{eqnarray}
for ATLAS and CMS, respectively.  
See App.~\ref{app:lims} for details on our updated combination.  There were also new searches for $h\rightarrow\gamma\gamma+ZZ$ at $\sqrt{S}=13.6$~TeV with $29$ fb$^{-1}$ from ATLAS~\cite{ATLAS:2023tnc}:
\begin{eqnarray}
\mu^{{\rm ATLAS},13.6~{\rm TeV}}_{\gamma\gamma,ZZ}=0.98\pm0.15.\label{eq:ATLASgamgamZZ}
\end{eqnarray}
Additionally, CMS~\cite{CMS:2023tfj} has a new measurement $h\rightarrow b\bar{b}$ in the vector boson fusion (VBF) production channel at $\sqrt{S}=13$~TeV with $90.8$~fb$^{-1}$:
\begin{eqnarray}
\mu^{VBF}_{bb,{\rm CMS}} = 1.01^{+0.55}_{-0.46}.\label{eq:CMSVBFbb}
\end{eqnarray}
The signal strengths in Eqs.~(\ref{eq:global},\ref{eq:ATLASgamgamZZ},\ref{eq:CMSVBFbb}) are used to find the 95\% CL upper limit on the mixing angle from precision Higgs measurements:
\begin{eqnarray}
|\sin\theta_1|\leq 0.29.
\end{eqnarray}
This value is consistent with current results in other literature~\cite{Bechtle:2020uwn,Bahl:2022igd,Robens:2022oue}.

In addition to Higgs signal strengths, direct searches for heavy scalars must be considered.  Indeed, these searches are often more constraining than the Higgs signal strengths~\cite{Adhikari:2020vqo,Robens:2022oue}. From Eq.~(\ref{eq:mixinglim}), we see that in the limit $|\theta_2|\ll 1$, the couplings between $h_2$ and fermions and gauge bosons are suppressed by $\sin\theta_1$.  As such, the production rate and partial widths into SM gauge bosons and fermions are suppressed by $\sin^2\theta_1$:
\begin{eqnarray}
\sigma(pp\rightarrow h_2)\approx \sin^2\theta_1 \sigma_{\rm SM}(pp\rightarrow h_2),\quad \Gamma(h_2\rightarrow f_{\rm SM})\approx \sin^2\theta_1 \Gamma_{\rm SM}(h_2\rightarrow f_{\rm SM}),
\end{eqnarray}
where $\sigma_{\rm SM}$ and $\Gamma_{\rm SM}$ indicate SM Higgs rates at the mass $m_2$, and $f_{\rm SM}$ are SM gauge boson and fermion final states.  If kinematically allowed $h_2$ may also decay into $h_1h_1$, $h_1h_3$, or $h_3h_3$.  Thus the total width is
\begin{eqnarray}
\Gamma_{\rm Tot}(h_2)\approx\sin^2\theta_1 \Gamma_{\rm SM}(h_2)+\Gamma(h_2\rightarrow h_1 h_1)+\Gamma(h_2\rightarrow h_1h_3)+\Gamma(h_2\rightarrow h_3\,h_3),
\end{eqnarray}
where $\Gamma_{\rm SM}(h_2)$ is the total SM-like width at the mass $m_2$.  From this it can be shown that the branching ratio of $h_2$ into SM final states is
\begin{eqnarray}
&&{\rm BR}(h_2\rightarrow f_{\rm SM})\approx\\
&&\quad\quad\quad\quad\quad{\rm BR}_{\rm SM}(h_2\rightarrow f_{\rm SM})\left[1-{\rm BR}(h_2\rightarrow h_1h_1)-{\rm BR}(h_2\rightarrow h_1h_3)-{\rm BR}(h_2\rightarrow h_3h_3)\right].\nonumber
\end{eqnarray}
Hence, in our fits we will treat
\begin{eqnarray}
m_2,\,\sin^2\theta_1,\,{\rm BR}(h_2\rightarrow h_1h_1),\,{\rm BR}(h_2\rightarrow h_1h_3),\,{\rm and}\,{\rm BR}(h_2\rightarrow h_3h_3)
\end{eqnarray}
as free parameters.  For all numerical results in this section, SM-like rates and branching ratios are set using the LHC Higgs Cross Section Working Group suggested values~\cite{deFlorian:2016spz}.

\begin{figure}[tb]
\subfigure[]{\includegraphics[width=0.45\textwidth,clip]{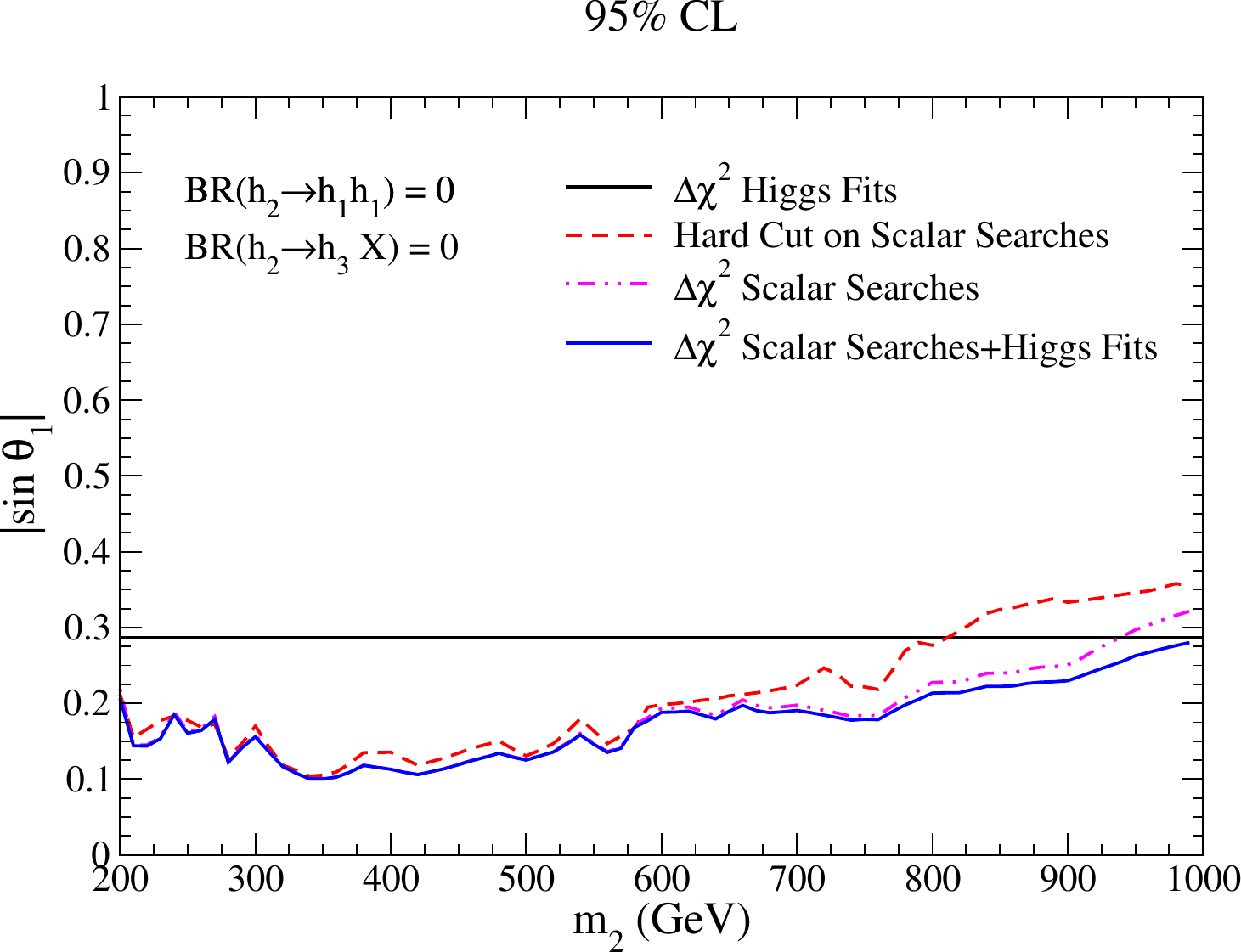}\label{fig:lims_BR0}}
\subfigure[]{\includegraphics[width=0.45\textwidth,clip]{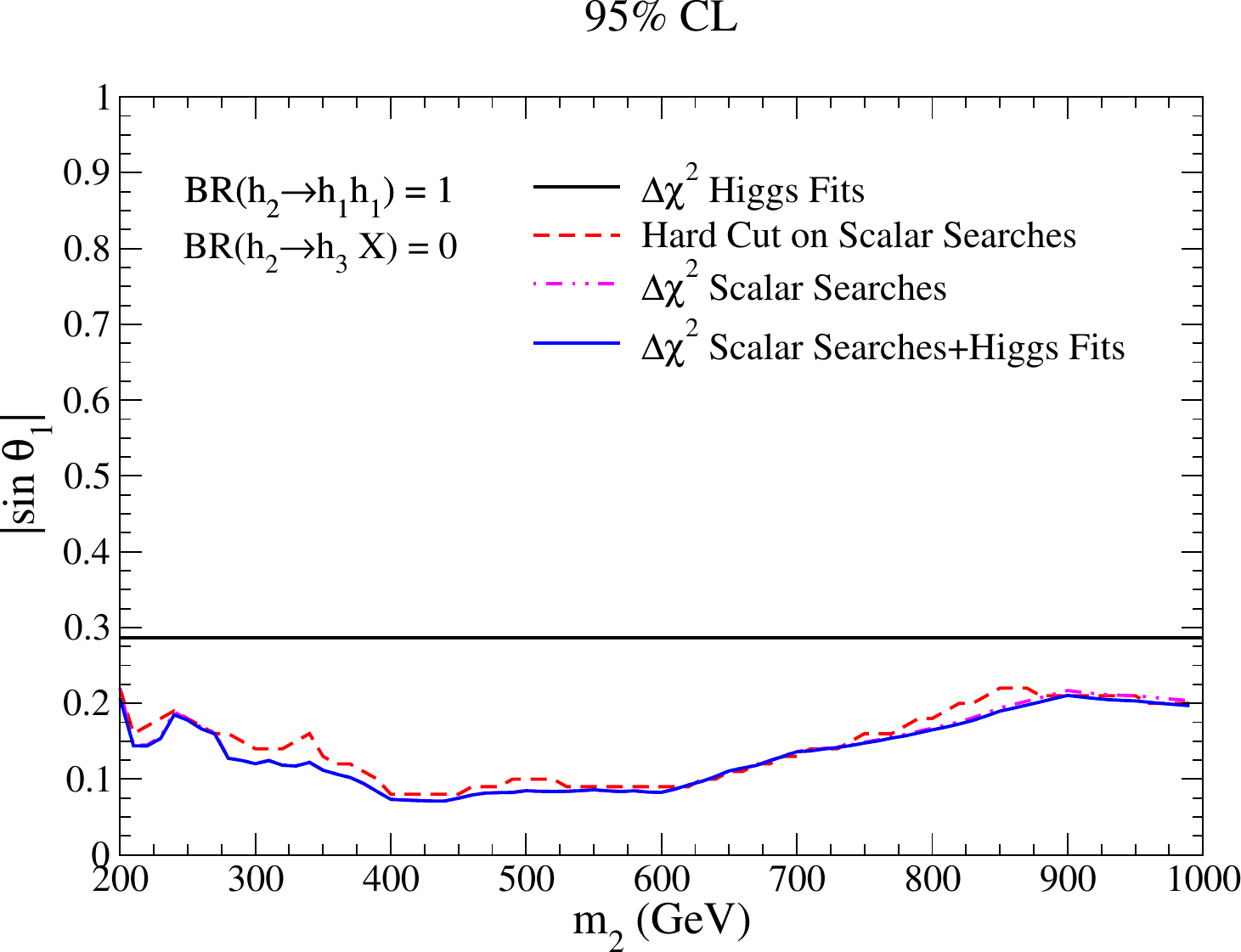}\label{fig:lims_BR1}}
\subfigure[]{\includegraphics[width=0.45\textwidth,clip]{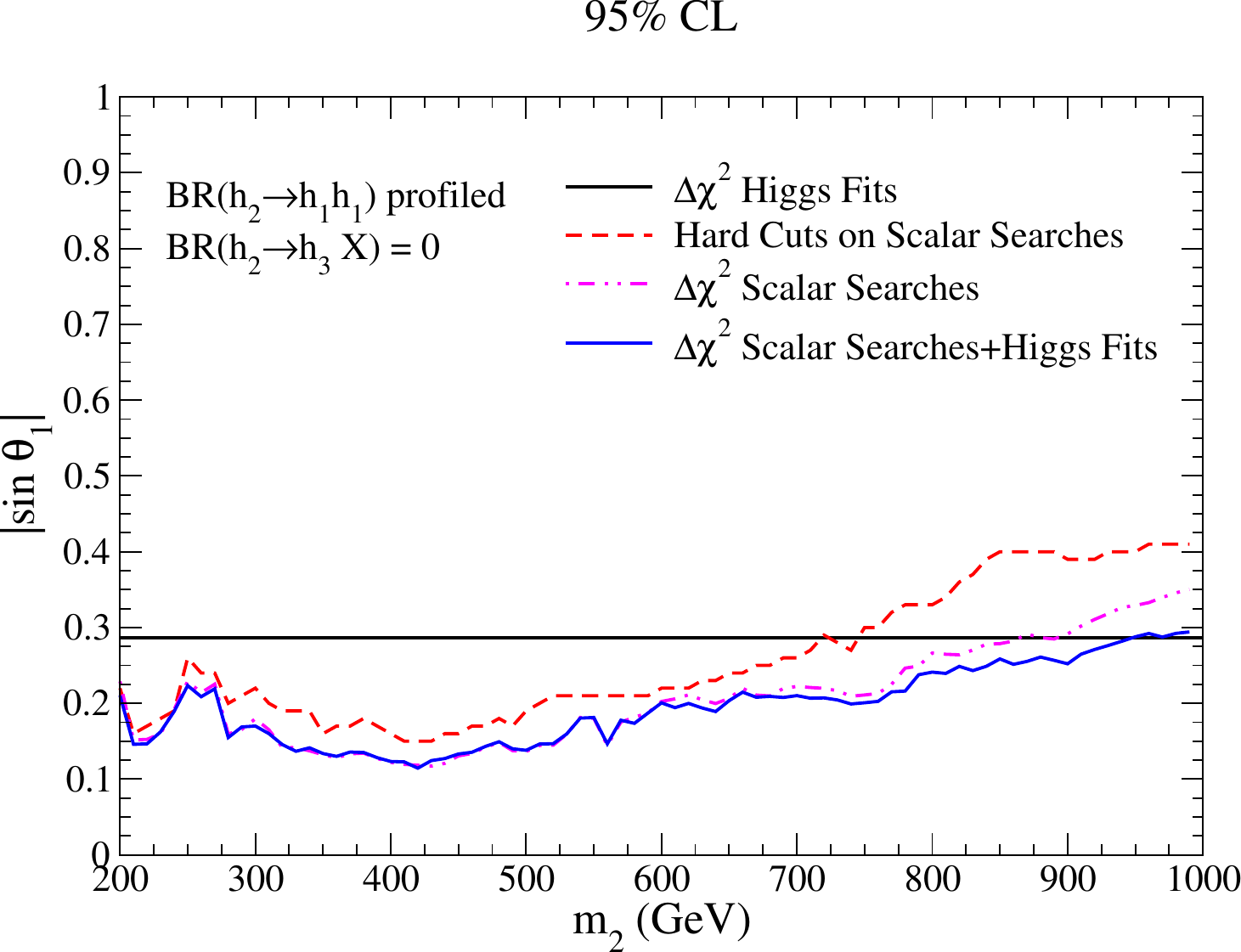}\label{fig:lims_BRFloat}}
\subfigure[]{\includegraphics[width=0.45\textwidth,clip]{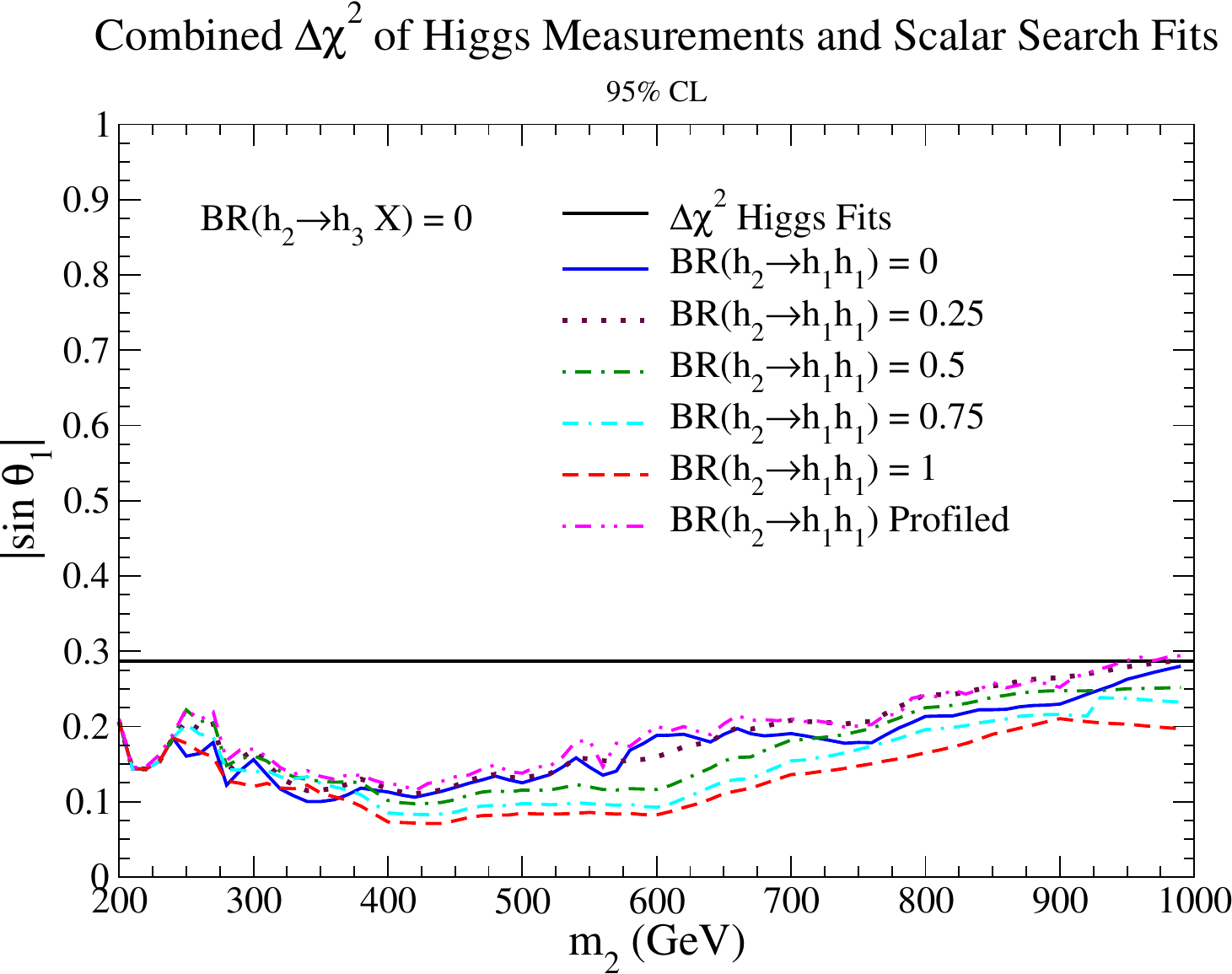}}
\caption{\label{fig:lims} The 95\% CL upper limits on $|\sin\theta_1|$ for (a) ${\rm BR}(h_2\rightarrow h_1h_1)=0$, (b) ${\rm BR}(h_2\rightarrow h_1h_1)=1$, (c) ${\rm BR}(h_2\rightarrow h_1h_1)$ profiled over, and (d) a variety of ${\rm BR}(h_2\rightarrow h_1h_1)$.  In all plots (black solid) the bounds from Higgs precision are shown.  In (a-c) the bounds resulting from the (red dashed) hard cuts on scalar searches, (magenta dot-dot-dashed) $\Delta \chi^2$ scalar search fits from Eq.~(\ref{eq:Chi2h2}), and (blue solid) $\Delta\chi^2$ combination of scalar searches and Higgs precision from Eq.~(\ref{eq:Chi2Tot}). The results in (d) are from the $\Delta\chi^2$ combination of scalar searches and Higgs precision. }
\end{figure}

 Traditionally, a parameter point is accepted if the predicted cross section is lower than all of the observed 95\% upper limits on cross sections reported in searches for new particles.  The point is rejected if the predicted cross section exceeds an observed cross section limit in any channel.  For example, see Refs.~\cite{Bechtle:2013xfa,Bechtle:2020uwn}.  We refer to this as the ``hard cut'' method.  The red dashed curves in Figs.~\ref{fig:lims}(a,b,c) show the resulting limits on $\sin\theta_1$ from utilizing the hard cut method.  The list of included experimental results are given in Table~\ref{tab:ScalSearch} in App.~\ref{app:lims}.  Since $h_2$ couples to SM fermions and gauge bosons like the SM Higgs except suppressed by $\sin\theta_1$, its production and decay are similar to a SM Higgs of the same mass, except when the additional decay to scalars is available.  Therefore, its dominate decays will be to gauge bosons or scalars and we only consider those experimental limits.

In Fig.~\ref{fig:lims} we always consider the case when ${\rm BR}(h_2\rightarrow h_1h_3)={\rm BR}(h_2\rightarrow h_2h_3)=0$.  If those branching ratios are nonzero, then direct searches for $h_2\rightarrow WW/ZZ/h_1h_1$ can be evaded by setting the branching ratios into the $h_1h_3$ and/or $h_3h_3$ final states to one.  In Fig.~\ref{fig:lims_BR0} we consider the case where ${\rm BR}(h_2\rightarrow h_1h_1)=0$.  As can be seen, the scalar resonance searches into $WW/ZZ$ are more constraining than Higgs signal strengths for $m_2\lesssim 800$~GeV in the hard cut scenario.  We also considered a scenario in which ${\rm BR}(h_2\rightarrow h_1h_1)=1$ and only searches for $h_2\rightarrow h_1h_1$ are relevant.  This is shown in Fig.~\ref{fig:lims_BR1}. These searches are always more constraining than the Higgs precision observables.  Finally, for the hard cut limits we also allowed ${\rm BR}(h_2\rightarrow h_1h_1)$ to float.  In this case, a parameter point was accepted if there exists a ${\rm BR}(h_2\rightarrow h_1h_1)$ such that it passes the hard cut method.  The results are shown in Fig.~\ref{fig:lims_BRFloat}.  Allowing the $h_2\rightarrow h_1h_1$ to float weakens the scalar search constraints which are now more stringent than precision Higgs only when $m_2\lesssim 700-725$~GeV. This can be understood by noting that for any $\sin\theta_1$ or $m_2$ this procedure will try to find a ${\rm BR}(h_2\rightarrow h_1h_1)$ that will evade constraints.  Hence, it weakens all direct search constraints.

While typically employed, the hard cut method does not allow for large fluctuations in individual channels that may be allowed when all searches are considered, nor does it account for strong constraints when signal cross sections may uniformly increase in terms of a model parameter.  Additionally, the hard cut method is not a true 95\% CL, which would rely on a likelihood function.  Once such a likelihood function is determined, the scalar searches can be consistently combined with Higgs precision measurements to obtain a 95\% CL upper bound on $|\sin\theta_1|$.  Following a previous proposal from a subset of the authors for combining scalar searches and Higgs precision data~\cite{Adhikari:2020vqo}, we start with a series of assumptions:
\begin{itemize}
\item All searches are consistent with SM backgrounds and have enough events to be Gaussian.  Then the search is essentially a measurement of SM backgrounds, and the reported 95\% CL upper limits on cross sections are 95\% uncertainty bands around the SM background measurement.
\item Assuming searches are consistent with SM predictions, the uncertainty on a measurement is well approximated by the expected 95\% CL limit.  
\item Any non-zero central value can be parameterized as the difference between the observed and expected 95\% CL as reported by the experiments.  If there is a downward fluctuation, we set the central value to zero and use the observed upper limit as the uncertainty.
\item We use the narrow width approximation, which is standard when applying search constraints~\cite{Bechtle:2011sb,Bechtle:2013wla,Bechtle:2020pkv}.  For heavy scalars, the interference between signal and background can be up to order $10\%$ in the resonance region~\cite{Kauer:2015hia,Dawson:2015haa,Greiner:2015ixr,Carena:2018vpt}.
\end{itemize}
With these assumptions, Ref.~\cite{Adhikari:2020vqo}  developed a channel-by-channel $\chi^2$ for heavy resonance searches:
\begin{eqnarray}
\left(\chi^{f}_{i,h_2}\right)^2=\begin{cases}
\displaystyle\left(\frac{\sigma_i(pp\rightarrow h_2){\rm BR}(h_2\rightarrow f)+\hat{\sigma}_{i,Exp}^f-\hat{\sigma}_{i,Obs}^f}{\hat{\sigma}_{i,Exp}^f/1.96}\right)^2 & {\rm if~} \hat{\sigma}_{i,Obs}^f\geq \hat{\sigma}_{i,Exp}^f\\
\displaystyle\left(\frac{\sigma_i(pp\rightarrow h_2){\rm BR}(h_2\rightarrow f)}{\hat{\sigma}_{i,Obs}^f/1.96}\right)^2&{\rm if~} \hat{\sigma}_{i,Obs}^f< \hat{\sigma}_{i,Exp}^f.
\end{cases}~\label{eq:Chi2Scalar}
\end{eqnarray}
where $\sigma_i(pp\rightarrow h_2)$ is the resonance production cross section from initial state $i$, ${\rm BR}(h_2\rightarrow f)$ is the branching ratio into final state $f$, and $\hat{\sigma}_{i,Exp}^f$ ($\hat{\sigma}_{i,Obs}^f$) is the expected (observed) 95\%CL upper limit on $\sigma(i\rightarrow h_2){\rm BR}(h_2\rightarrow f)$.  It can be checked that, for an individual search, this $\chi^2$ reproduces the limit $\sigma(i\rightarrow h_2){\rm BR}(h_2\rightarrow f)<\hat{\sigma}_{i,Obs}^f$, consistent with the hard cut method.  With this definition, all heavy scalar search channels can be consistently combined:
\begin{eqnarray}
\chi^2_{h_2}=\sum_{i,f}\left(\chi^{f}_{i,h_2}\right)^2.\label{eq:Chi2h2}
\end{eqnarray}
Additionally heavy scalar searches and Higgs precision measurements can be consistently combined into a global $\chi^2$:
\begin{eqnarray}
\chi^2_{\rm Tot}=\chi^2_{h_1}+\chi^2_{h_2}.\label{eq:Chi2Tot}
\end{eqnarray}

The magenta dot-dot-dashed lines in Figs.~\ref{fig:lims}(a,b,c) show the 95\% CL upper limits on $|\sin\theta_1|$ that result from fitting scalar searches according to Eq.~(\ref{eq:Chi2h2}).  For all limits we have assumed ${\rm BR}(h_2\rightarrow h_1h_3)={\rm BR}(h_2\rightarrow h_3h_3)=0$.  In (a) we set ${\rm BR}(h_2\rightarrow h_1h_1)=0$, in (b) ${\rm BR}(h_2\rightarrow h_1h_1)=1$, and in (c) we profile over ${\rm BR}(h_2\rightarrow h_1h_1)$.  As can be clearly seen, the results of Eq.~(\ref{eq:Chi2h2}) have similar behavior to and are consistently stronger than the hard cuts.  This can be understood by noting that since all production cross sections of $h_2$ are proportional to $\sin^2\theta_1$, any increase in $|\sin\theta_1|$ essentially increases all signal rates in all channels uniformly increasing the $\chi^2$ value.  Accordingly, we would more-or-less expect to see a global upward fluctuation in the channels listed in Tab.~\ref{tab:ScalSearch} if the scalar resonance exists.  Such a global increase should result in a stronger constraint.  Our $\chi^2$ method accounts for this, while the hard cut method does not.

The result from combining the scalar searches and Higgs precision data from Eq.~(\ref{eq:Chi2Tot}) are shown as blue solid lines in Figs.~\ref{fig:lims} (a,b,c).  For ${\rm BR}(h_2\rightarrow h_1h_1)=1$, the combined $\chi^2_{\rm Tot}$ fit gives very similar results as the scalar search $\chi^2_{h_2}$.  For the results with ${\rm BR}(h_2\rightarrow h_1h_1)=0$ or ${\rm BR}(h_2\rightarrow h_1h_1)$ profiled over, combining the Higgs precision constraints with the scalar searches strengthens the bounds in the large mass region: $m_2\gtrsim 800$~GeV.  Indeed, although the scalar search constraints may get weaker than the Higgs precision at very high mass, the combined Higgs and scalar search results stay more or similarly stringent as the Higgs measurement results.

We compare the effects of many different hypotheses for ${\rm BR}(h_2\rightarrow h_1h_1)$ in Fig.~\ref{fig:lims}(d).  These results are from the combination of the Higgs precision and scalar searches in Eq.~(\ref{eq:Chi2Tot}).  As can be seen, for all mass ranges under consideration, either ${\rm BR}(h_2\rightarrow h_1h_1)=0$ or $1$ gives the most stringent constraints.  For mass $m_2\gtrsim 350$~GeV, ${\rm BR}(h_2\rightarrow h_1h_1)=1$ is always the most constraining.  Comparing the results for the hard cuts give similar conclusions.

There are also constraints on the singlet mixing angle from electroweak precision observables~\cite{Lopez-Val:2014jva,deBlas:2016ojx,Dawson:2017jja,Ilnicka:2018def}.  For small $\theta_2$, fits to the oblique parameters~\cite{Dawson:2017jja,Ilnicka:2018def} are always weaker than the most stringent constraints in Fig.~\ref{fig:lims}.  The strongest electroweak precision constraint comes from fitting the $W$-mass individually~\cite{Lopez-Val:2014jva}.  This is most relevant in the high mass region where the $W$-mass limit is $|\sin\theta_1|\lesssim 0.20-0.21$ for $m_2\gtrsim 800$~GeV~\cite{Lopez-Val:2014jva,Dawson:2017jja}.  Again, the most stringent constraints in Fig.~\ref{fig:lims} are as strong or stronger than the $W$-mass limit.  Hence, for the rest of the paper we consider constraints on $\sin\theta_1$ from direct searches and Higgs precision data only.

\begin{figure}[tb]
\subfigure[]{\includegraphics[width=0.45\textwidth,clip]{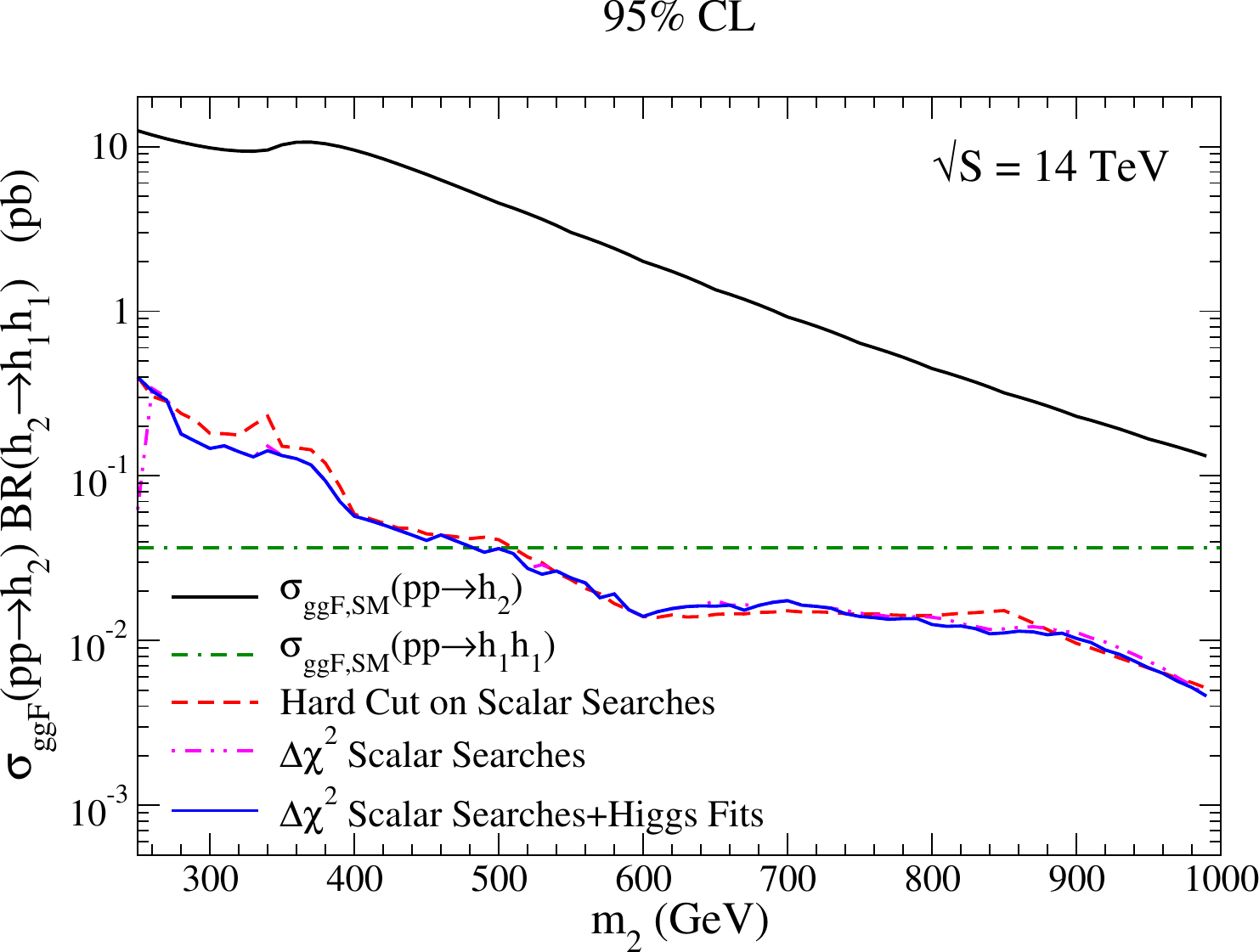}}
\subfigure[]{\includegraphics[width=0.45\textwidth,clip]{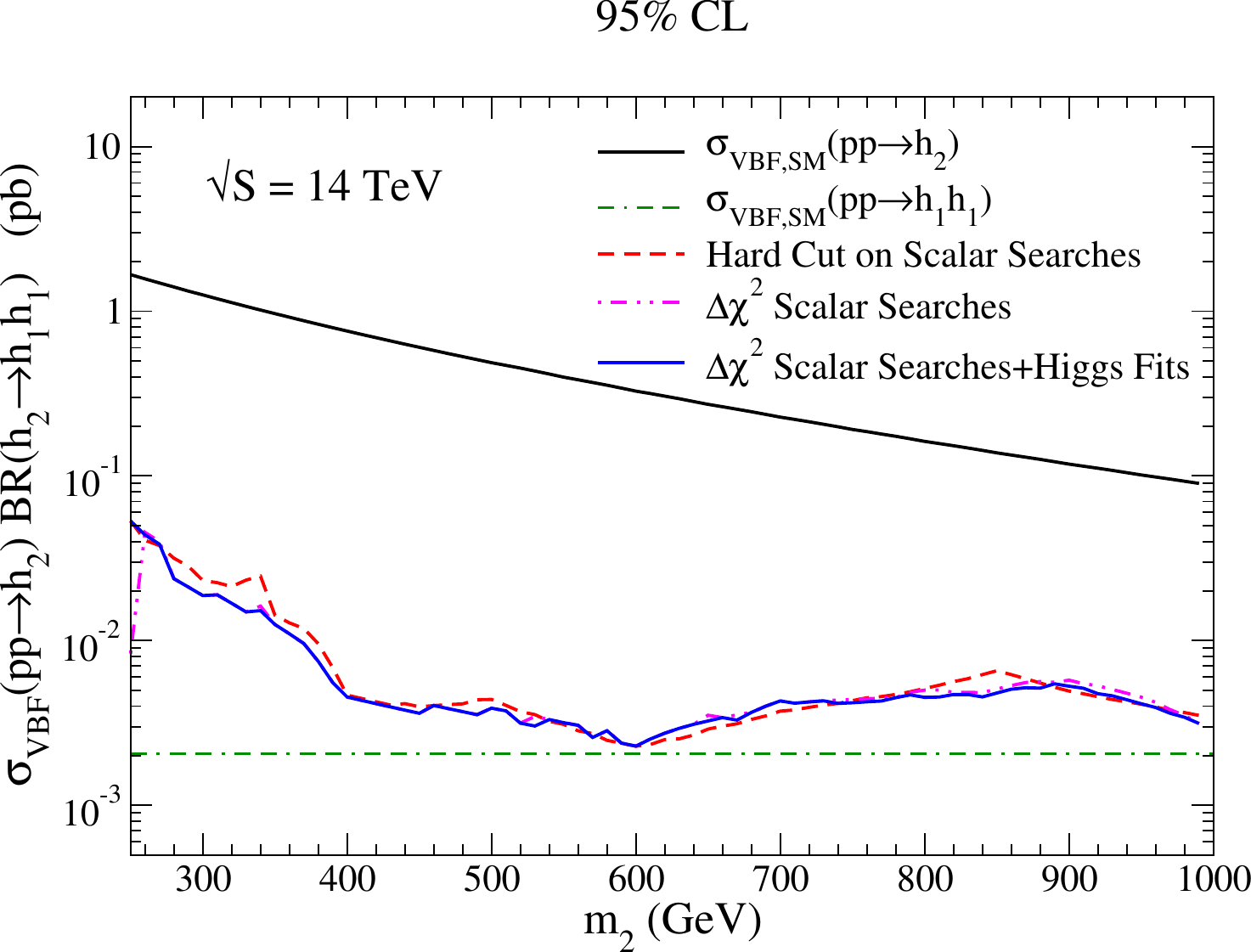}}
\caption{\label{fig:DiHiggslim} Maximum allowed resonant production rates for $pp\rightarrow h_2\rightarrow h_1h_1$ in (a) gluon fusion and (b) VBF channels for (red dashed) hard cut bounds, (magenta dot-dot-dashed) $\Delta\chi^2$ bounds on scalar searches from Eq.~(\ref{eq:Chi2h2}), and (blue solid) $\Delta\chi^2$ bounds on scalar searches and Higgs precision from Eq.~(\ref{eq:Chi2Tot}).  For comparison, SM predictions for (black solid) $h_2$ production and (green dot-dash-dashed) $h_1h_1$ production are shown.}
\end{figure}

In Fig.~\ref{fig:DiHiggslim} we show 95\% CL upper limits on the cross section for production and decay $pp\rightarrow h_2\rightarrow h_1h_1$ in both the (a) gluon fusion and (b) VBF channels.  These bounds are found by performing a simultaneous fit to $\sin\theta_1$ and $\sin^2\theta_1{\rm BR}(h_2\rightarrow h_1h_1$) with ${\rm BR}(h_2\rightarrow h_1h_3)={\rm BR}(h_2\rightarrow h_3h_3)=0$.  For the $\Delta\chi^2$ fits, $\sin\theta_1$ is profiled over.  For the hard cut, a value of $\sin^2\theta_1 {\rm BR}(h_2\rightarrow h_1h_1)$ is accepted if there exists a $\sin\theta_1$ that passes all cuts and Higgs precision data.  For comparison, the SM single $h_2$ and double $h_1h_1$ production rates are shown.  The $\Delta\chi^2$ and hard cut bounds produce very similar limits on $pp\rightarrow h_2\rightarrow h_1h_1$.  For $m_2\lesssim500$~GeV, the resonant gluon fusion di-Higgs rate can still be above the SM di-Higgs rate rate, up to an order of magnitude larger.  For VBF, the allowed rate is above the SM VBF di-Higgs rate for all masses shown.  The resonant VBF rate can be up to a factor of 20 larger than the SM VBF di-Higgs prediction.

\section{Benchmark Points}
\label{sec:benchmark}

We now present the main results of this paper: finding the maximum rates for $pp\rightarrow h_2\rightarrow h_ih_j$, $i,j=1,3$, considering all theoretical and experimental constraints.  As mentioned in the Introduction, we consider several possible scenarios for different Higgs precision measurements and direct scalar searches: 
\begin{itemize}
\item S1: We apply the current limits, described in Section~\ref{sec:Const}, on $\sin\theta_1$ to produce expected maximum rates for the HL-LHC. The most constraining limits on $\sin\theta_1$ from Fig.~\ref{fig:lims} are used.  Results from both the hard cut method for scalar searches with Higgs precision data constraints and $\chi^2_{\rm Tot}$ fits from Eq.~(\ref{eq:Chi2Tot}) are considered separately.  In practice, the strongest constraints on $\sin\theta_1$ are found by considering the smallest maximum value of $|\sin\theta_1|$ from the ${\rm BR}(h_2\rightarrow h_1h_1)=0$ and ${\rm BR}(h_2\rightarrow h_1h_1)=1$ cases.
\item S2: We use the European Strategy Report (ESR) projected constraints on $|\sin\theta_1|$~\cite{EuropeanStrategyforParticlePhysicsPreparatoryGroup:2019qin} for HL-LHC from both Higgs precision measurements and direct scalar searches.  
\item S3: We use the ESR projected constraints on $|\sin\theta_1|$~\cite{EuropeanStrategyforParticlePhysicsPreparatoryGroup:2019qin} for the HL-LHC and FCC-ee.
\item S4: Finally, we use the ESR projected constraints on $|\sin\theta_1|$~ \cite{EuropeanStrategyforParticlePhysicsPreparatoryGroup:2019qin} for HL-LHC and ILC500.
\end{itemize}
Scenario S1 is intended to find maximum di-scalar resonance rates for the HL-LHC.  Hence, the $h_2$ mass range $m_i+m_j\leq m_2\leq 1$~TeV is used for S1, where $m_i$ and $m_j$ are the masses of the $h_2$ decay products.  Scenarios S2-S4 are used to project maximum rates for future $pp$ machines, such as the 100 TeV FCC-hh.  In these scenarios we consider a mass range of $m_i+m_j\leq m_2\leq12$~TeV.  For scenarios S2-S4, we normalize to the maximum rates to the SM Higgs production rate at the mass of $h_2$.  Hence, they are very generic and can be applied to many future collider proposals and many  single $h_2$ production modes.

\begin{figure}
\begin{center}
\subfigure[]{\includegraphics[width=0.45\textwidth,clip]{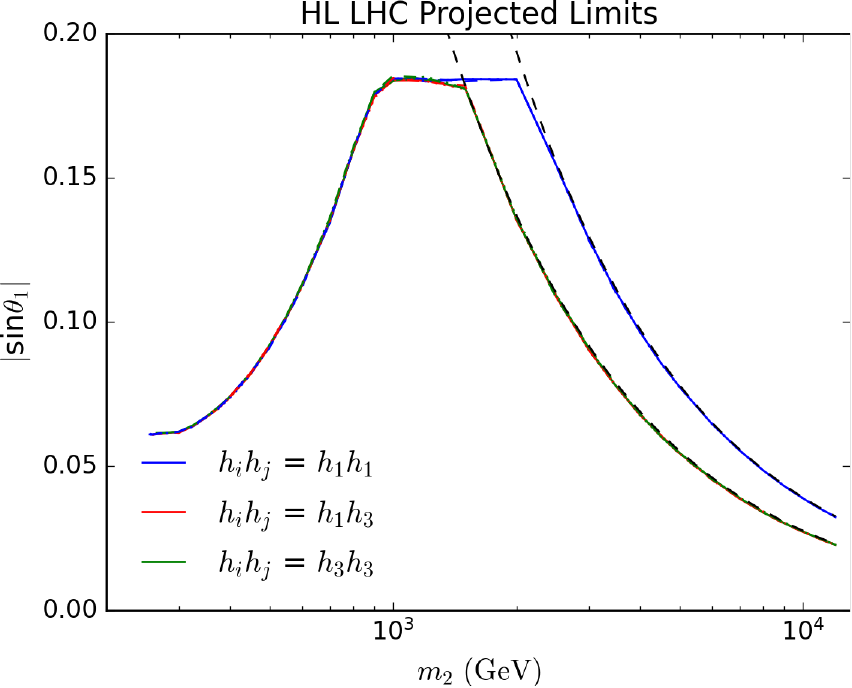}}
\subfigure[]{\includegraphics[width=0.45\textwidth,clip]{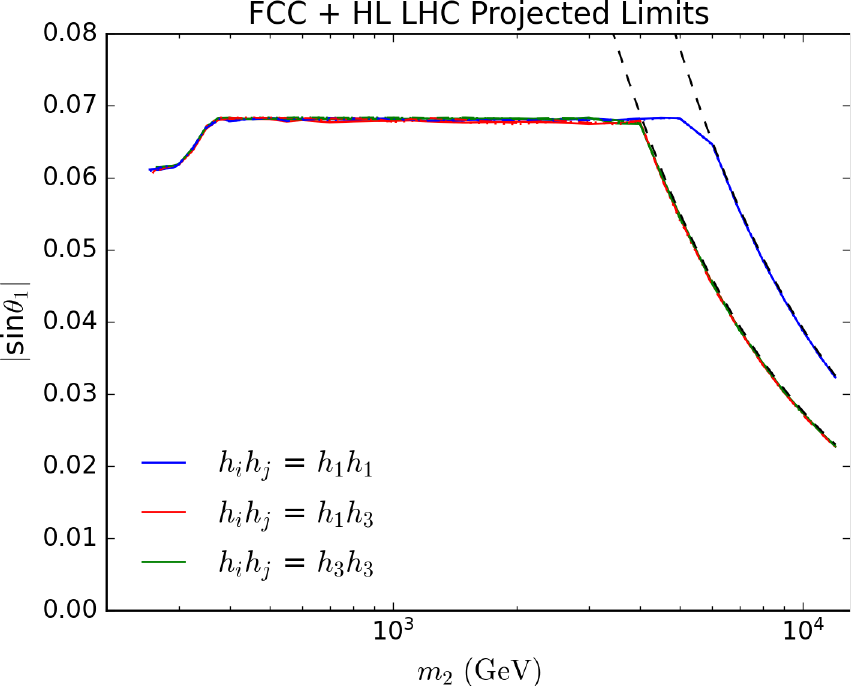}}
\subfigure[]{\includegraphics[width=0.45\textwidth,clip]{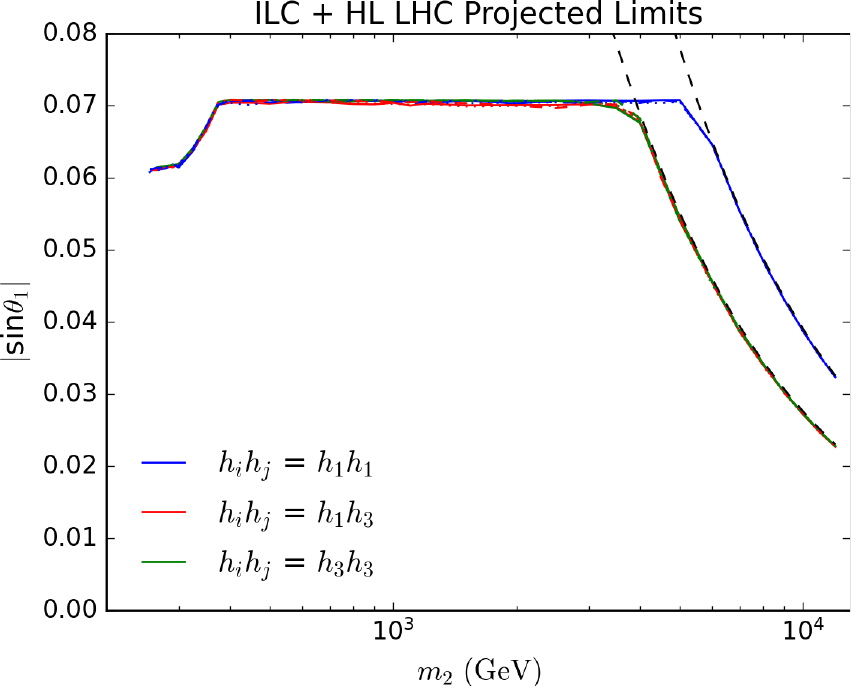}}
\end{center}
\caption{\label{fig:sinthetaFuture} Values of $|\sin\theta_1|$ found from maximizing rates for scenarios (a) S2, (b) S3, and (c) S4. Blue is for $h_2\rightarrow h_1h_1$, red for $h_2\rightarrow h_1h_3$, and green for $h_2\rightarrow h_3h_3$. Solid is for $m_3=130$~GeV, dotted for $m_3=200$~GeV, and dashed for $m_3=270$~GeV.  The dashed black lines show analytical upper bounds on $|\sin\theta_1|$ from maximizing the $h_ih_j$ with the narrow width constraint in Eq.~(\ref{eq:sthMaxh3h3}) for $h_1h_3/h_3h_3$ production and Eq.~(\ref{eq:sthMaxh1h1}) for $h_1h_1$ production, with $\kappa=0.1$.}
\end{figure}

For completeness, in Fig.~\ref{fig:sinthetaFuture} we show the values of $|\sin\theta_1|$ used for scenarios S2-S4.  The low mass region follows constraints from direct searches for scalar production at the HL-LHC.  As $m_2$ increases, the values of $|\sin\theta_1|$ plateau and flatten out.  The flat regions come from the respective constraints from precision Higgs measurements, which are independent of $m_2$.  After the flat region, there is a deep decrease in the value of $|\sin\theta_1|$.  This occurs from a combination of maximizing rates and requiring a narrow width, which will be discussed below.

While a range of $m_2$ are considered, only certain values of the mass of $h_3$ are used.  These masses are chosen to open new decay channels for $h_3$ and possibly new phenomenology for the resonance searches.  Assuming $\theta_2$ is negligible but non-zero, $h_3$ will have partial widths into SM gauge bosons and fermions:
\begin{eqnarray}
\Gamma(h_3\rightarrow f_{\rm SM})=\sin^2\theta_1\sin^2\theta_2\,\Gamma_{\rm SM}(h_3\rightarrow f_{\rm SM}),\label{eq:h3width}
\end{eqnarray}
where $\Gamma_{\rm SM}(h_3\rightarrow f_{\rm SM})$ is the partial width of a SM-like Higgs boson with mass $m_3$ into SM fermions or gauge bosons collectively denoted as $f_{\rm SM}$.  A negligible $\sin\theta_2$ will not alter the calculations in the previous section, nor the maximized rates presented here.    

Three possible masses of $h_3$ are used: $m_3=130$~GeV, $m_3=200$~GeV, and $m_3=270$~GeV.  For the masses $130$ and $200$~GeV, $h_3$ decays like a SM Higgs at those masses.  That is, for $m_3=130$~GeV the dominant decays are $h_3\rightarrow b\bar{b}$ and $h_3\rightarrow W^\pm W^{\mp,*}$.  Hence, we could have multi-$b$ and multi-$W$ resonances from $h_2$ decays.  The largest branching ratio through $h_3$ would be
\begin{eqnarray}
h_2\rightarrow h_1h_3/h_3h_3\rightarrow 2b\,2\overline{b}.
\end{eqnarray}
With $m_3=200$ GeV, the on-shell decays $h_3\rightarrow W^\pm W^\mp$ and $h_3\rightarrow ZZ$ open up with the decay into $W$s dominating.  In which case, the predominant decays of $h_2$ into $h_3$ would be 
\begin{eqnarray}
h_2\rightarrow h_1h_3\rightarrow b\overline{b}W^\pm W^\mp\quad{\rm and}\quad h_2\rightarrow h_3h_3\rightarrow 2W^\pm\,2W^\mp.
\end{eqnarray}
The final mass point $m_3=270$~GeV is above the di-Higgs threshold. Hence, it is possible to have a resonance of three or four SM-like Higgs bosons:
\begin{eqnarray}
h_2\rightarrow h_1h_3\rightarrow 3\,h_1\quad{\rm and}\quad h_2\rightarrow h_3h_3\rightarrow 4\,h_1.
\end{eqnarray}
Whether or not these are the dominant branching ratios depends on the Higgs trilinear coupling $h_1-h_1-h_3$ which in turn depends on the parameter point that maximizes the $h_2\rightarrow h_1h_3$ or $h_2\rightarrow h_3h_3$ production rates. As we show below, for all benchmark points, the branching ratio of $h_3\rightarrow h_1h_1$ is at or near one when kinematically allowed. 

For the benchmark masses $m_3=130$ and $200$~GeV, there are possible decays of $h_3$ into off-shell $h_1$:  $h_3 \rightarrow h_1^* h_1^*$ and $h_3 \rightarrow h_1 h_1^*$. In principle, these decays could compete with the on-shell decays $h_3\rightarrow bb$ and $h_3\rightarrow WW$. The branching ratios of $h_3$ into off-shell $h_1$ will depend on the relevant trilinear: $\lambda_{113}$.  Upper limits on $\lambda_{113}$ can be found using the bounds on scalar potential parameters in App.~\ref{app:CubicBounds}. As discussed below, there is a subtlety on whether or not the rate of $h_2\rightarrow h_1h_3$ or $h_2\rightarrow h_3h_3$ is being maximized.  As seen in Eqs.~(\ref{eq:trilinear123},\ref{eq:trilinear113}), in the small angle limit, the leading behavior of $\lambda_{113}$ is the same as $\lambda_{123}$.  Hence, if $h_2\rightarrow h_1h_3$ is being maximized, the relevant upper bound on the $h_1-h_1-h_3$ coupling is
\begin{eqnarray}
|\lambda_{113}|&\lesssim& 4\sqrt{\frac{2\pi}{3}}m_1|\sin\theta_1|\left|\cos\theta_1\sqrt{1+\frac{m_2^2-m_1^2}{m_1^2}\sin^2\theta_1}\right.\label{eq:lam113Max1}\\
&&\left.+\sqrt{2}\sin\theta_1\frac{m_2}{m_1}\left(\sqrt{1+\frac{m^2_3}{m^2_2}-\frac{m_2^2-m_1^2}{m_2^2}\sin^2\theta_1}+\frac{m_3}{2\,m_2}\right)\right|+\mathcal{O}(\sin\theta_2)\nonumber
\end{eqnarray}
The rate for $h_2\rightarrow h_3h_3$ is maximized when the rate for $h_2\rightarrow h_1h_3$ is minimized, i.e. $\lambda_{123}=0$.  This corresponds to the first term in Eq.~(\ref{eq:trilinear113}) being zero.  In this case the relevant upper bound on $\lambda_{113}$ 
\begin{eqnarray}
|\lambda_{113}|&\lesssim &|\sin\theta_1\sin\theta_2|\left|\frac{2\,m_1^2+m_3^2}{v_{\rm EW}}\cos^2\theta_1+4\sqrt{\frac{2\pi}{3}}m_1\left[\left(1-\frac{3}{2}\sin^2\theta_1\right)\sqrt{1+\frac{m_2^2-m_1^2}{m_1^2}\sin^2\theta_1}\right.\right.\nonumber\\
&&\left.\left.+\frac{3}{2^{3/2}}\sin2\theta_1\frac{m_2}{m_1}\sqrt{1-\frac{m_2^2-m_1^2}{m_2^2}\sin^2\theta_1}\right]\right|+\mathcal{O}(\sin^2\theta_2)\label{eq:lam113Max2}
\end{eqnarray}
For a benchmark point we use $\sin\theta_2=0.01$.  For either of the upper bounds in Eqs.~(\ref{eq:lam113Max1},\ref{eq:lam113Max2}), we find that at $m_3=130$~GeV the branching ratio of $h_3\rightarrow h_1^*h_1^*$ is at least four orders of magnitude smaller than ${\rm BR}(h_3\rightarrow b\bar{b})$. At $m_3=200$ GeV, when the bound in Eq.~(\ref{eq:lam113Max1}) is saturated the branching ratio of $h_3\rightarrow h_1h_1^*$ is a factor of $2-20$ smaller than $h_3\rightarrow WW$, while for the bound in Eq.~(\ref{eq:lam113Max2}) $h_3\rightarrow h_1h_1^*$ is at least four orders of magnitude smaller than $h_3\rightarrow WW$. Hence, $h_3\rightarrow b\bar{b}$ is the dominate decay at $m_3=130$~GeV and $h_3\rightarrow WW$ dominates at $m_3=200$~GeV.  Note that $h_3$ couplings to SM fermions and gauge bosons are suppressed by the combination $\sin\theta_1\sin\theta_2$. The bound in Eq.~(\ref{eq:lam113Max1}) is suppressed by only $\sin\theta_1$.  Hence, whether $h_3\rightarrow h_1h_1^*$ or $h_3\rightarrow WW$ are dominant at $m_3=200$~GeV can be sensitive to the choice of $\sin\theta_2$ when $h_2\rightarrow h_1h_3$ is being maximized.  At $m_3=130$~GeV, $h_3\rightarrow bb$ is expected to always be dominant due to the very large phase space and $h_1$ coupling constant suppression in $h_3\rightarrow h_1^*h_1^*$.  Finally, the bound in Eq.~(\ref{eq:lam113Max2}) is suppress by $\sin\theta_1\sin\theta_2$ and all results derived from that bound is robust against different choices of $\sin\theta_2$. 

We now discuss the results for maximizing $h_ih_j$ resonant production under the various scenarios.  After these results, we will comment on the decays of $h_3$ and the possibility of having three or four SM-like Higgs boson signals.

\subsection{Results}

\subsubsection{Scenario S1: HL-LHC projections}
\begin{figure}[tb]
\subfigure[]{\includegraphics[height=0.27\textheight,clip]{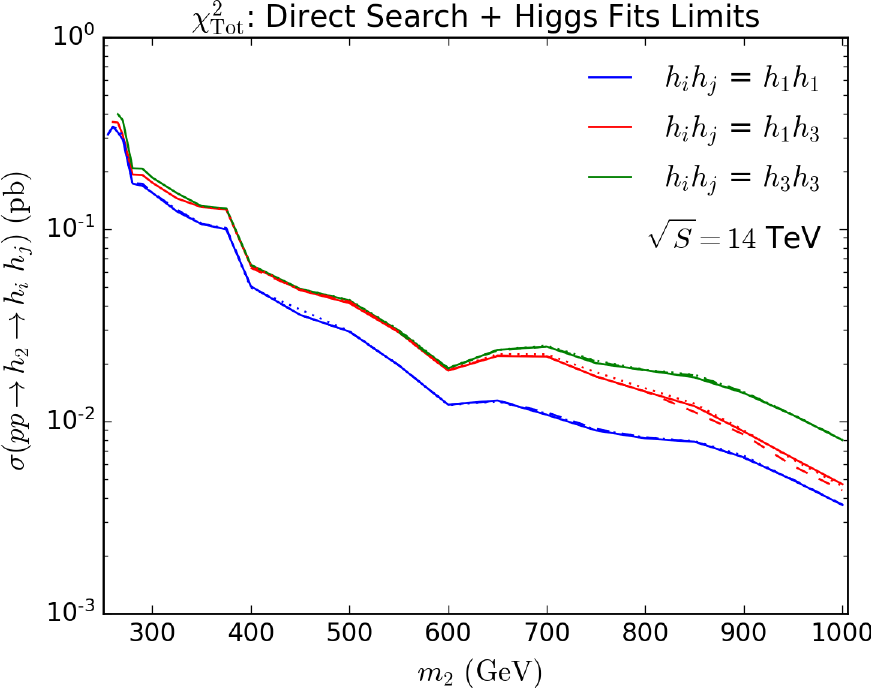}}
\subfigure[]{\includegraphics[height=0.27\textheight,clip]{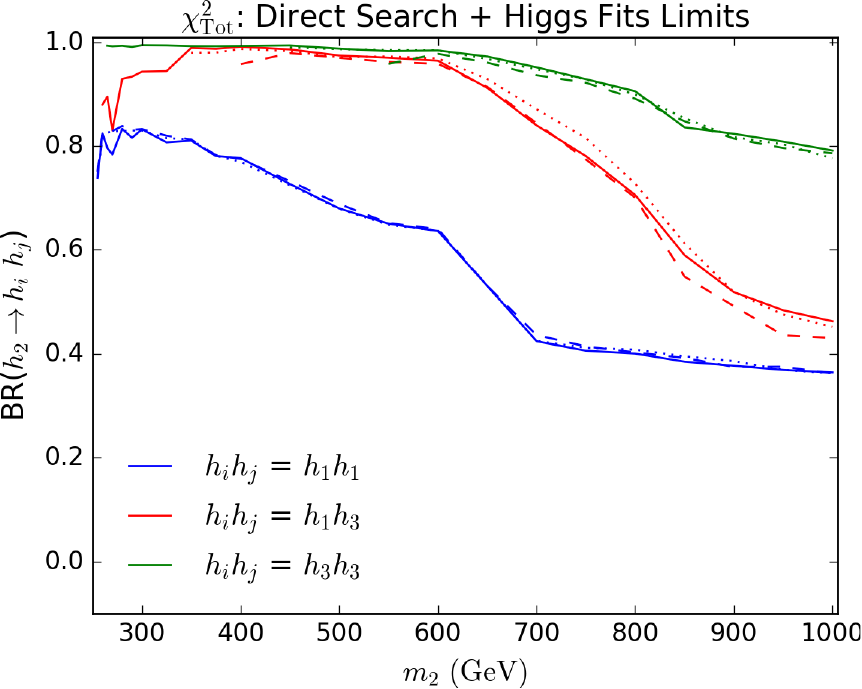}}
\subfigure[]{\includegraphics[height=0.27\textheight,clip]{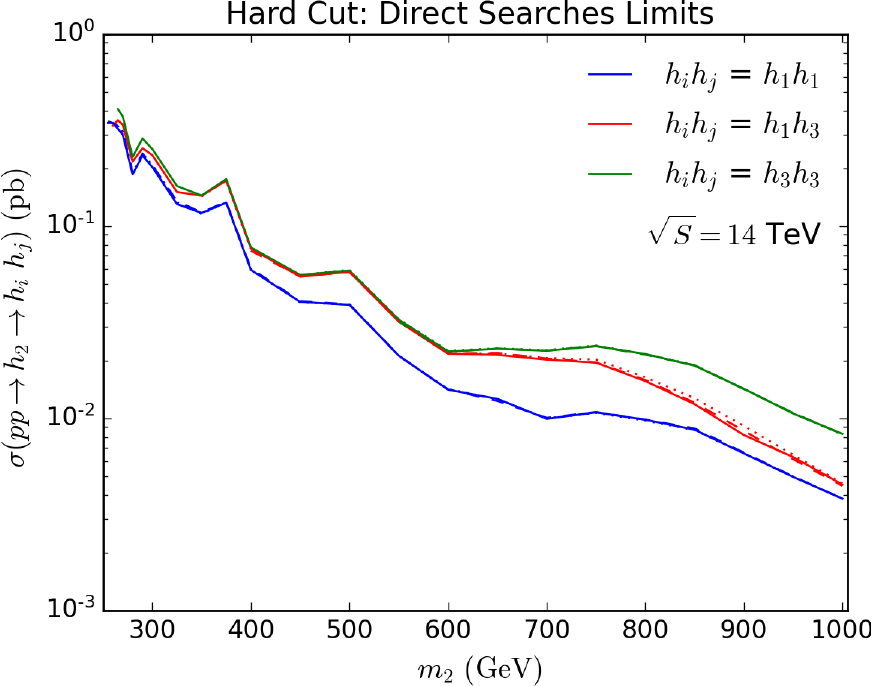}}
\subfigure[]{\includegraphics[height=0.27\textheight,clip]{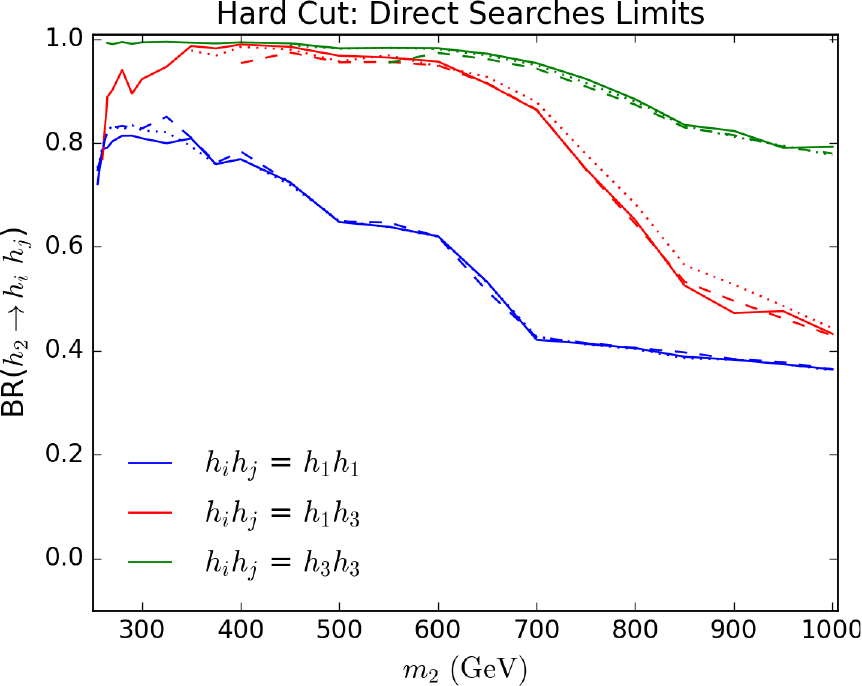}}
\caption{\label{fig:S1plots}Benchmarks for HL-LHC from scenario S1, showing (a,c) maximum production rates for $pp\rightarrow h_2\rightarrow h_ih_j$ with the gluon fusion and VBF modes added together at $\sqrt{S}=14$~TeV and (b,d) the corresponding branching ratios to different di-scalar states. Results from using the bounds from $\Delta\chi^2_{\rm Tot}$ fits in Eq.~(\ref{eq:Chi2Tot}) are shown in (a,b) and from the hard cut method in (c,d). Blue is for $h_2\rightarrow h_1h_1$, red for $h_2\rightarrow h_1h_3$, and green for $h_2\rightarrow h_3h_3$. Solid is for $m_3=130$~GeV, dotted for $m_3=200$~GeV, and dashed for $m_3=270$~GeV. }
\end{figure}

In figure \ref{fig:S1plots} we show the upper bounds on (a,c) the maximum production rate for $pp\rightarrow h_2\rightarrow h_ih_j$ and (b,d) the corresponding branching ratios of $h_2 \rightarrow h_i h_j$, $i,j=1,3$, for scenario S1. Figure~\ref{fig:S1plots}(a,b) use constraints from a $\chi^2_{\rm Tot}$ fit and (c,d) from the hard cut method for scalar searches with precision Higgs constraints. 
The maximum production rates and branching ratios are largely independent of the choice of $m_3$ once kinematically allowed.

For $m_2\lesssim 600$~GeV, the maximum branching
ratios for $h_2\rightarrow h_1h_3$ and $h_2\rightarrow h_3h_3$ are near one, with the final state $h_3h_3$ slightly larger.  For $h_2\rightarrow h_1h_1$ the maximum branching ratio is between $0.6$ and $0.8$. As can be seen in Eqs.~(\ref{eq:trilinear112},-\ref{eq:trilinear233}), the values of the $\lambda_{123}$ and $\lambda_{233}$ trilinear couplings have much more freedom from the choices of potential parameters than $\lambda_{112}$.  Indeed, the ratio of the $h_2\rightarrow h_1h_1$ and $h_2\rightarrow ZZ$ partial widths are
\begin{eqnarray}
\frac{\Gamma(h_2\rightarrow h_1h_1)}{\Gamma(h_2\rightarrow ZZ)}=1+\mathcal{O}\left(\sin^2\theta_1,\frac{v_{\rm EW}^2}{m_2^2},\sin\theta_1\frac{v_{\rm EW}}{m_2}\right).
\end{eqnarray}
Hence, the partial width of $h_2\rightarrow h_1h_1$ cannot be too far from $h_2\rightarrow ZZ$, limiting how large the branching ratio of $h_2\rightarrow h_1h_1$ can be. 

Once the mass of $h_2$ surpasses 600 GeV, all branching ratios decrease.  The partial widths of $h_2$ into the massive SM gauge bosons $W^\pm W^\mp$ and $ZZ$ grow like the cubic power of $m_2$, whereas the partial widths into $h_1h_3$ and $h_3h_3$ do not.  Additionally, the branching ratio of $h_2\rightarrow h_1h_1$ approaches the Goldstone boson equivalence theorem limit of ${\rm BR}(h_2\rightarrow h_1h_1)\approx0.25$.  Hence, the maximal branching ratio into $h_ih_j$ decreases in general.  It is worth noting that the maximum branching ratio of the asymmetric decay $h_2\rightarrow h_1h_3$ decreases more quickly than the symmetric decay of $h_2\rightarrow h_3h_3$.  In the small angle limit $|\theta_1|\ll 1$, the leading term of the trilinear $\lambda_{123}$ depends on trilinear couplings while $\lambda_{233}$ depends on quartic couplings.  As we show in Appendix~\ref{app:TrilinearBounds}, the combination of requiring the global minimum be the SM vacuum, perturbative unitarity, and a narrow width places an upper bound on $\lambda_{123}$ that is more stringent than the upper bound on $\lambda_{233}$.  Indeed, as shown in Eq.~(\ref{eq:h2h1h3Lim}), these constraints force the branching ratio of $h_2\rightarrow h_1h_3$ to approach zero as the mass of $h_2$ increases.  While these derived constraints may not be the most stringent, they do help explain qualitatively why the branching ratio of $h_2\rightarrow h_1h_3$ decreases more quickly than the other di-scalar modes as $m_2$ increases.

\subsubsection{Scenarios S2-S4: Future Collider Projections}
\begin{figure}[tb]
\subfigure[]{\includegraphics[height=0.27\textheight,clip]{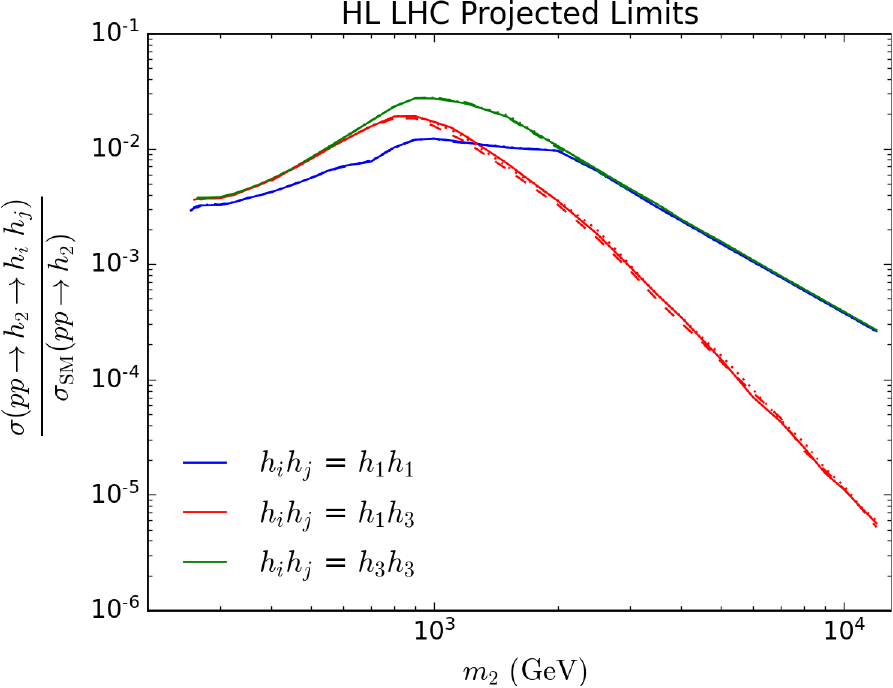}}\hspace{0.1in}{\includegraphics[height=0.27\textheight,clip]{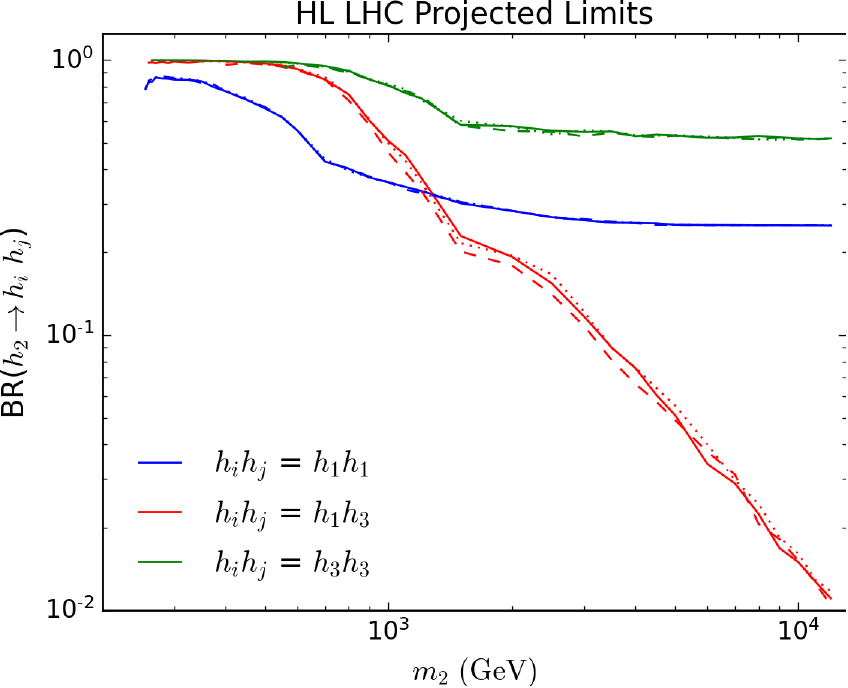}}

\caption{\label{fig:S2plots}Benchmarks for future colliders, constrained by projections for HL-LHC in scenario S2, showing (a) the maximum production times branching ratio scaled to the SM-like Higgs production rate and (b) the corresponding branching ratios. Blue is for $h_2\rightarrow h_1h_1$, red for $h_2\rightarrow h_1h_3$, and green for $h_2\rightarrow h_3h_3$. Solid is for $m_3=130$~GeV, dotted for $m_3=200$~GeV, and dashed for $m_3=270$~GeV. }
\end{figure}

\begin{figure}[tb]
\subfigure[]{\includegraphics[height=0.27\textheight,clip]{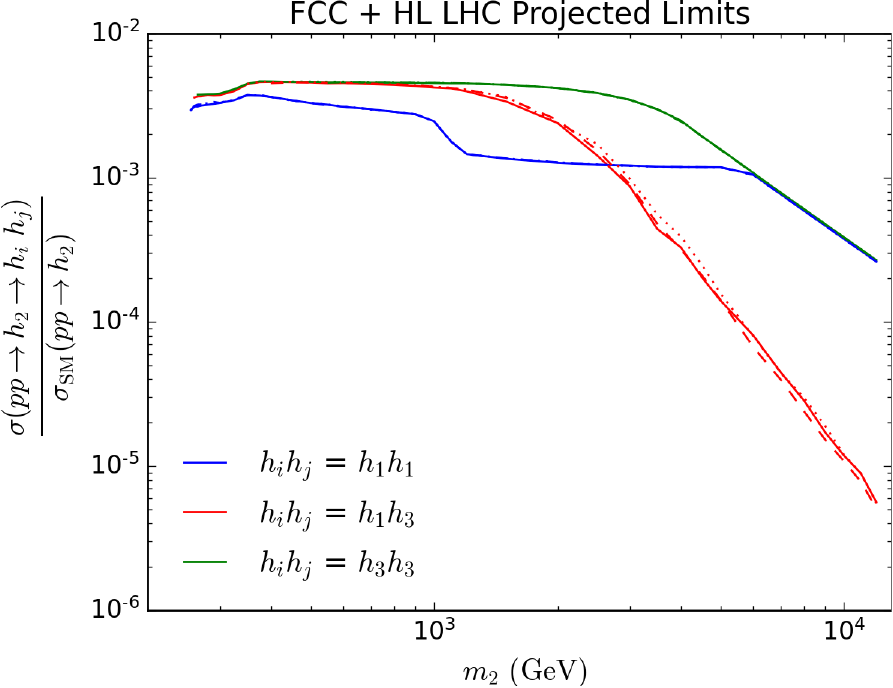}}\hspace{0.1in}
\subfigure[]{\includegraphics[height=0.27\textheight,clip]{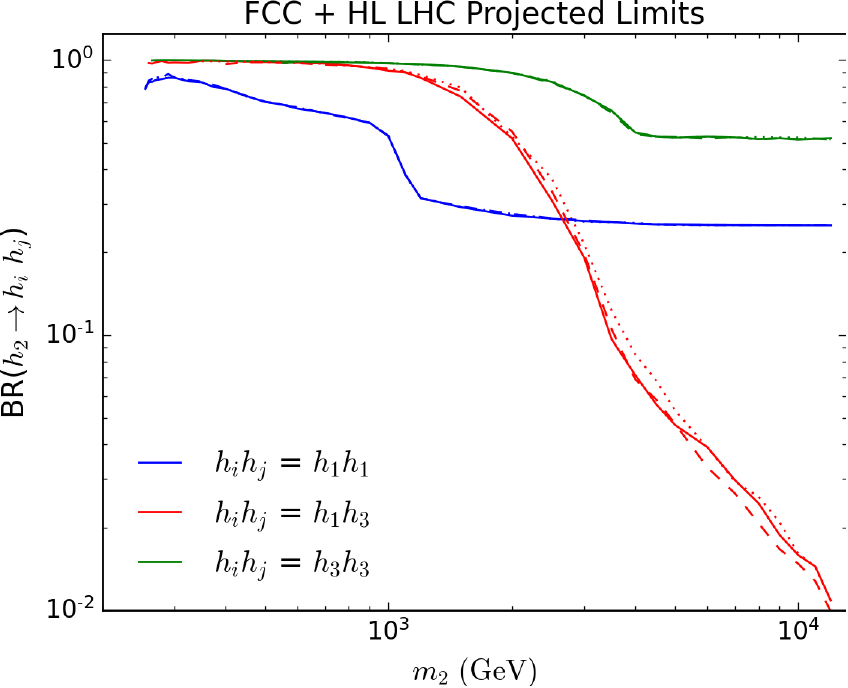}}
\caption{\label{fig:S3plots}Benchmarks for future colliders, constrained by projections for HL-LHC and FCC-ee in scenario S3, showing (a) the maximum production times branching ratio scaled to the SM-like Higgs production rate and (b) the corresponding branching ratios. Blue is for $h_2\rightarrow h_1h_1$, red for $h_2\rightarrow h_1h_3$, and green for $h_2\rightarrow h_3h_3$. Solid is for $m_3=130$~GeV, dotted for $m_3=200$~GeV, and dashed for $m_3=270$~GeV.}
\end{figure}

\begin{figure}[tb]
\subfigure[]{\includegraphics[height=0.27\textheight,clip]{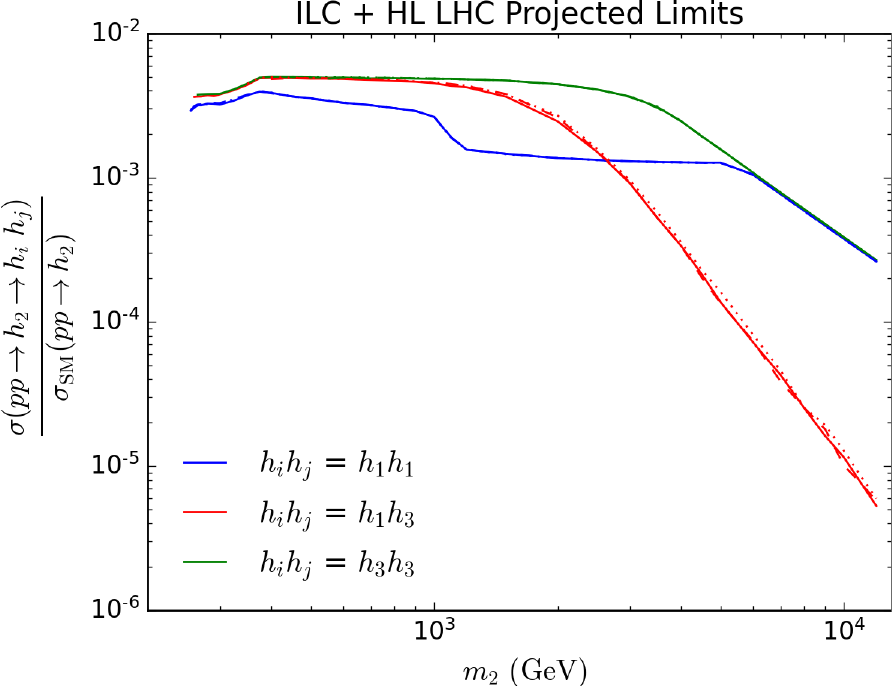}}\hspace{0.1in}
\subfigure[]{\includegraphics[height=0.27\textheight,clip]{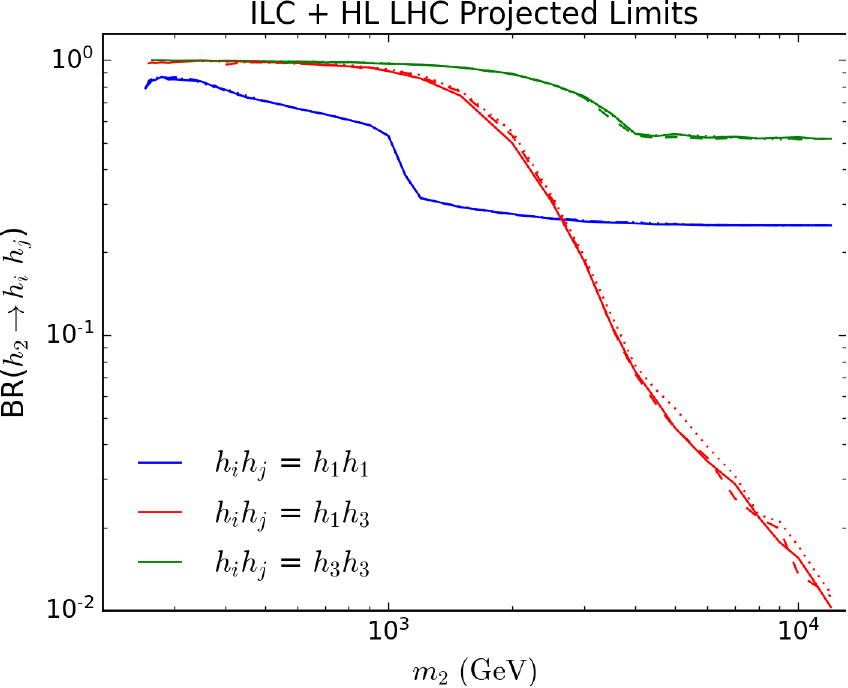}}

\caption{\label{fig:S4plots}Benchmarks for future colliders, constrained by projections for HL-LHC and ILC500 in scenario S4, showing (a) the maximum production times branching ratio scaled to the SM-like Higgs production rate and (b) the corresponding branching ratios. Blue is for $h_2\rightarrow h_1h_1$, red for $h_2\rightarrow h_1h_3$, and green for $h_2\rightarrow h_3h_3$. Solid is for $m_3=130$~GeV, dotted for $m_3=200$~GeV, and dashed for $m_3=270$~GeV}
\end{figure}

 We show (a) the maximum allowed production rates of $pp\rightarrow h_2 \rightarrow h_i h_j$ as well as (b) the corresponding branching ratios of $h_2\rightarrow h_ih_j$, $i,j=1,3$, in Figs. \ref{fig:S2plots},~\ref{fig:S3plots}, and~\ref{fig:S4plots} for scenarios S2, S3, and S4, respectively.  The production rate is normalized to the SM Higgs production rate at mass $m_2$, $\sigma_{\rm SM}(pp\rightarrow h_2)$.   There are many interesting features in these scenarios.  For $h_2$ masses below $1$~TeV, the relative behavior of the different di-scalar production modes are similar to those for the HL-LHC scenario S1, as discussed above.  Above 1 TeV, the maximum ${\rm BR}(h_2\rightarrow h_1h_3)$ still approaches zero quickly due to global minimum, perturbativity, and narrow width constraints causing $\lambda_{123}\rightarrow 0$ as $m_2$ becomes large, as discussed for scenario S1.  The maximum branching ratio of $h_2\rightarrow h_1h_1$ approaches one quarter in the multi-TeV range due to the Goldstone boson equivalence theorem. This is due to the possibility of choosing parameters consistent with theoretical constraints such that $\lambda_{123}=\lambda_{233}=0$ and $\lambda_{112}\neq 0$.  Hence, when $m_2$ is in the multi-TeV regime, the Goldstone boson equivalence theorem is the dominant effect and
 \begin{eqnarray}
 \Gamma(h_2\rightarrow W^\pm W^\mp)\approx 2\,\Gamma(h_2\rightarrow ZZ)\approx  2\,\Gamma(h_2\rightarrow h_1h_1). 
 \end{eqnarray}
 The fermionic partial widths only grow linearly with $m_2$ while the bosonic widths grow as $m_2^3$.  Hence, $h_2\rightarrow f\bar{f}$, where $f$ are SM fermions, can be neglected and ${\rm BR}(h_2\rightarrow h_1h_1)\approx 0.25$.

Two particularly surprising things occur when $m_2$ is in the multi-Tev range and above: the branching ratio of $h_2\rightarrow h_3h_3$ at its maximum rate approaches 0.5 and the maximum rates of $h_2\rightarrow h_1h_1$ and $h_2\rightarrow h_3h_3$ approach the same value.  To understand ${\rm BR}(h_2\rightarrow h_3h_3)\approx 0.5$, note that we are maximizing the rate
\begin{eqnarray}
\sigma(pp\rightarrow h_2\rightarrow h_3h_3)/\sigma_{\rm SM}(pp\rightarrow h_2)\approx\sin^2\theta_1 {\rm BR}(h_2\rightarrow h_3h_3),\label{eq:rate}
\end{eqnarray}
where we have used the narrow width approximation.  Assuming the rate $pp\rightarrow h_2\rightarrow h_3h_3$ is maximized then ${\rm BR}(h_2\rightarrow h_1h_3)=0$ and the fermionic branching ratios are negligible.  Equation~(\ref{eq:rate}) can be rewritten as
\begin{eqnarray}
\sin^2\theta_1 {\rm BR}(h_2\rightarrow h_3h_3)&=&\sin^2\theta_1\left(1-{\rm BR}(h_2\rightarrow W^\pm W^\mp)-{\rm BR}(h_2\rightarrow ZZ)-{\rm BR}(h_2\rightarrow h_1h_1)\right)\nonumber\\
&\approx& \sin^2\theta_1\left(1-\frac{4}{3}\sin^2\theta_1\frac{\Gamma_{\rm SM}(h_2)}{\Gamma_{\rm Tot}(h_2)}\right),\label{eq:rate1}
\end{eqnarray}
where in the last step the Goldstone boson equivalence theorem is used and $\Gamma_{\rm SM}(h_2)$ is the total width of a SM Higgs at the mass of $h_2$.  Since the widths into $W^\pm W^\mp$, $ZZ$, and $h_1h_1$ are fixed by the SM results and the Goldstone boson equivalence theorem, the total rate into $h_3h_3$ should be maximum when the narrow width bound on the total width is saturated:
\begin{eqnarray}
\kappa\,m_2\geq \Gamma_{\rm Tot}(h_2),
\end{eqnarray}
where for now we have chosen an arbitrary fraction $\kappa$ of the mass $m_2$ for generality.  Setting $\Gamma_{\rm Tot}(h_2)=\kappa\,m_2$, Eq.~(\ref{eq:rate1}) can be maximized with respect to $\sin\theta_1$ to find the value of $\sin\theta_1$ at the maximum rate, the branching ratio of $h_2\rightarrow h_3h_3$ at the maximum rate, and the maximum rate:
\begin{eqnarray}
\sin^2\theta_{1,{\rm max}}&=&\frac{3}{8}\frac{\kappa\,m_2}{\Gamma_{\rm SM}(h_2)},\label{eq:sthMaxh3h3}\\
{\rm BR}_{\rm max}(h_2\rightarrow h_3h_3)&=&\frac{1}{2},\label{eq:BRMaxh3h3}\\
\sin^2\theta_{1,{\rm max}}{\rm BR}_{\rm max}(h_2\rightarrow h_3h_3)&=&\frac{3}{16}\frac{\kappa\,m_2}{\Gamma_{\rm SM}(h_2)},\label{eq:RateMaxh3h3}
\end{eqnarray}
where the subscript ``${\rm max}$'' indicates values evaluated at the maximum rate.  This explains why ${\rm BR}(h_2\rightarrow h_3h_3)\approx 0.5$ when the rate is maximized. Additionally, the dashed line in Fig.~\ref{fig:sinthetaFuture} that agrees with the $|\sin\theta_1|$ that maximizes the $h_1h_3$ and $h_3h_3$ rates in the multi-TeV $h_2$ mass range corresponds to Eq.~(\ref{eq:sthMaxh3h3}) with $\kappa=0.1$.

A similar calculation can be made for $pp\rightarrow h_1h_3$, which explains why $|\sin\theta_1|$ at the maximum rate agrees for $h_1h_3$ and $h_3h_3$ production [see Fig.~\ref{fig:sinthetaFuture}].  As discussed previously, theoretical constraints force ${\rm BR}(h_2\rightarrow h_1h_3)\rightarrow 0$ as $m_2$ becomes large.  Hence, branching ratios and rates in Eqs.~(\ref{eq:BRMaxh3h3},\ref{eq:RateMaxh3h3}) cannot be saturated.  There are constraints on $\lambda_{233}$ from global minimum, perturbative unitarity, and narrow width as shown in Eq.~(\ref{eq:lam233Bound}).  However, these are significantly weaker than for $\lambda_{123}$, and the $pp\rightarrow h_2\rightarrow h_3h_3$ can saturate the bounds in Eqs.~(\ref{eq:BRMaxh3h3},\ref{eq:RateMaxh3h3}).

For $h_2\rightarrow h_1h_1$, in the multi-TeV range, the maximum rate is
\begin{eqnarray}
\sin^2\theta_{1,{\rm max}} {\rm BR}_{\rm max}(h_2\rightarrow h_1h_1)\approx\frac{1}{4}\sin^2\theta_{1,{\rm max}},
\end{eqnarray}
where the Goldstone boson equivalence theorem result of ${\rm BR}(h_2\rightarrow h_1h_1)=0.25$ was used.  
That is, the rate is maximized at the maximum allowed $\sin^2\theta_{1,{\rm max}}$.  In the multi-TeV range, the narrow width requirement places significant bounds on the value of $\sin\theta_{1,{\rm max}}$ since the width into bosons grows like the cubic power of $m_2$.  Assuming $\Gamma(h_2\rightarrow h_1h_3)=\Gamma(h_2\rightarrow h_3h_3)=0$ when maximizing $h_2\rightarrow h_1h_1$, we have
\begin{eqnarray}
\kappa\,m_2\gtrsim \Gamma_{\rm Tot}(h_2)\approx \frac{4}{3}\sin^2\theta_1\,\Gamma_{\rm Tot,SM}(h_2).
\end{eqnarray}
The maximum $\sin^2\theta_1$ and $h_2\rightarrow h_1h_1$ rates can be solved for:
\begin{eqnarray}
\sin^2\theta_{1,{\rm max}}&=&\frac{3}{4}\frac{\kappa\,m_2}{\Gamma_{\rm SM}(h_2)}\label{eq:sthh1h1}\label{eq:sthMaxh1h1}\\
\sin^2\theta_{1,{\rm max}}{\rm BR}_{\rm max}(h_2\rightarrow h_1h_1)&=&\frac{3}{16}\frac{\kappa\,m_2}{\Gamma_{\rm SM}(h_2)}.\label{eq:RateMaxh1h1}
\end{eqnarray}
The maximum rates for $pp\rightarrow h_2\rightarrow h_3h_3$ in Eq.~(\ref{eq:RateMaxh3h3}) and $pp\rightarrow h_2\rightarrow h_1h_1$ in Eq.~(\ref{eq:RateMaxh1h1}) are the same, explaining why these maximum rates agree in scenarios S2-S4 as shown in Figs.~\ref{fig:S2plots}(a), \ref{fig:S3plots}(a), and \ref{fig:S4plots}(a).  Additionally, for $\kappa=0.1$, Eq.~(\ref{eq:sthMaxh1h1}) is the dashed line that converges to the $|\sin\theta_1|$ value when maximizing $pp\rightarrow h_2\rightarrow h_1h_1$ in the multi-TeV region.

\subsection{$h_3\rightarrow h_1h_1$ decays and multi-Higgs signals}

\begin{figure*}[p]
\begin{center}
\subfigure[]{\includegraphics[height=0.25\textheight,clip]{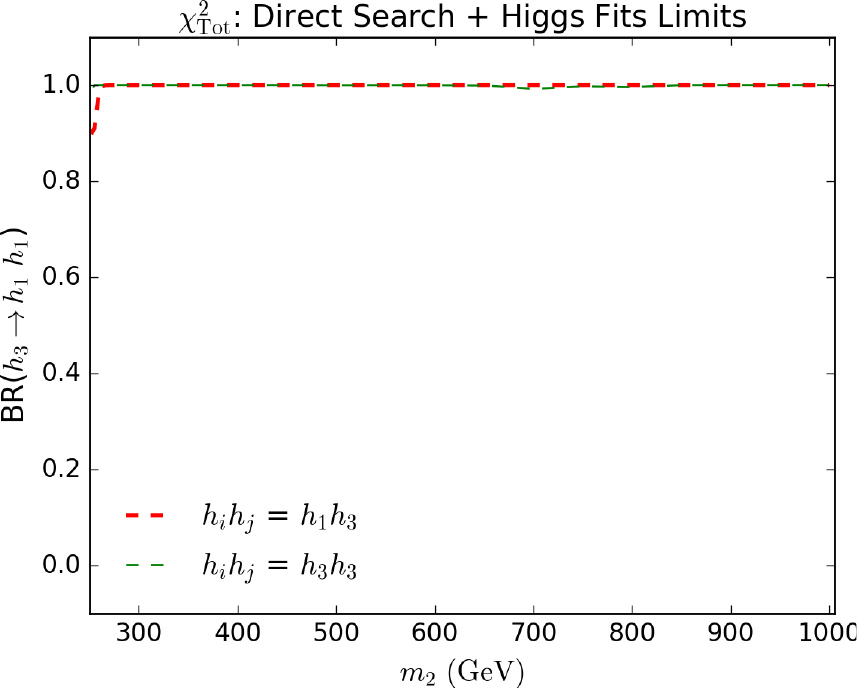}}\hspace{0.1in}
\subfigure[]{\includegraphics[height=0.25\textheight,clip]{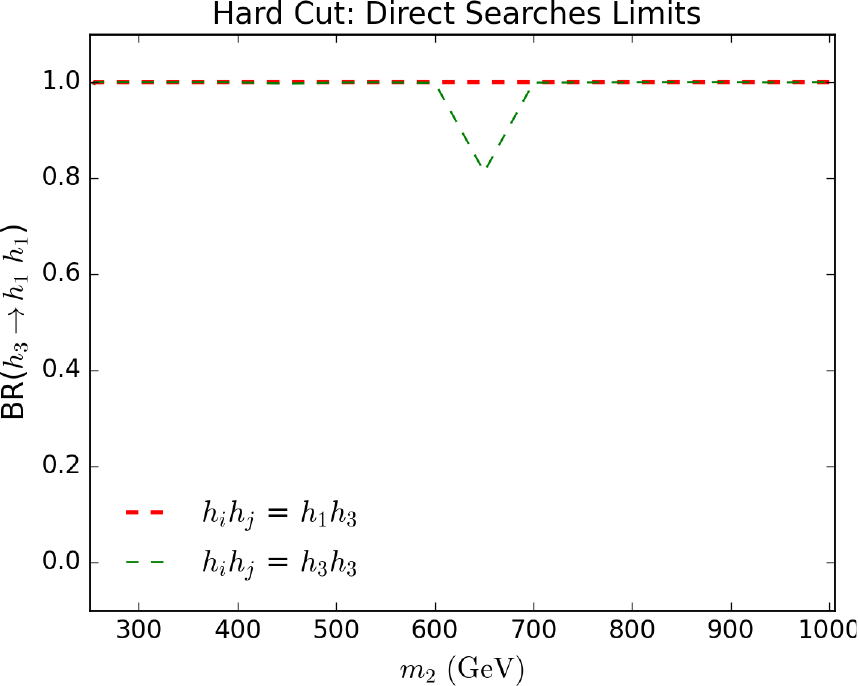}}
\subfigure[]{\includegraphics[height=0.25\textheight,clip]{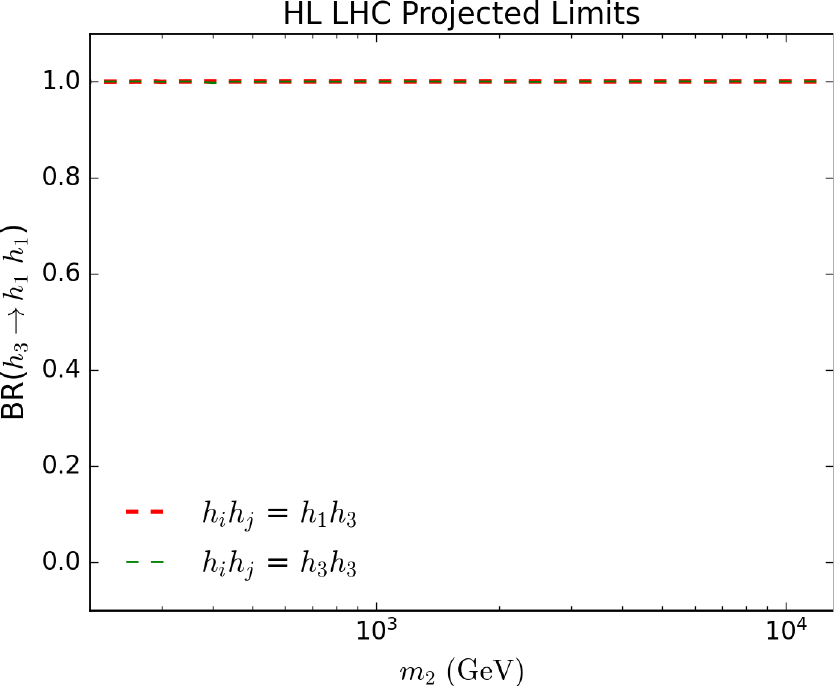}}\hspace{0.1in}
\subfigure[]{\includegraphics[height=0.25\textheight,clip]{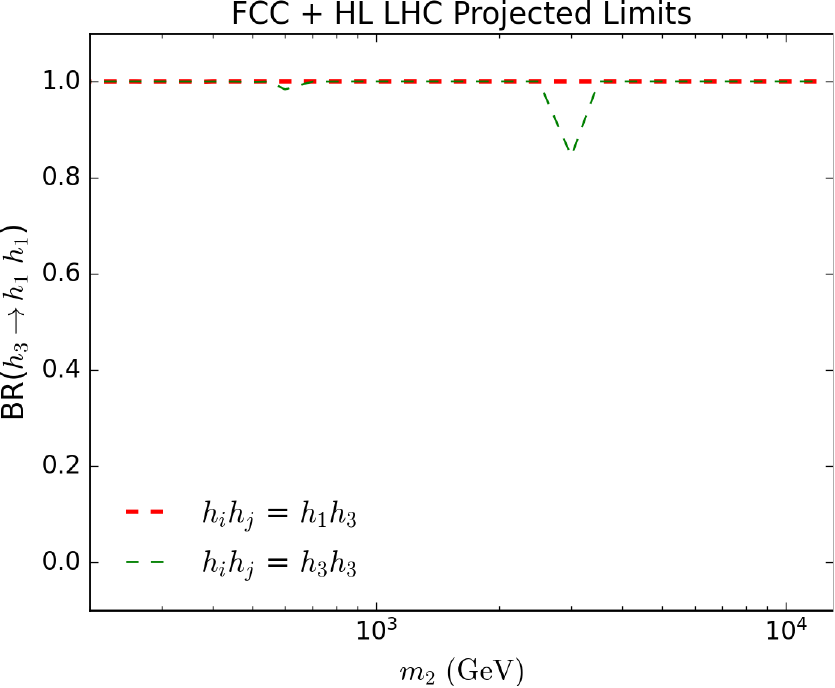}}
\subfigure[]{\includegraphics[height=0.25\textheight,clip]{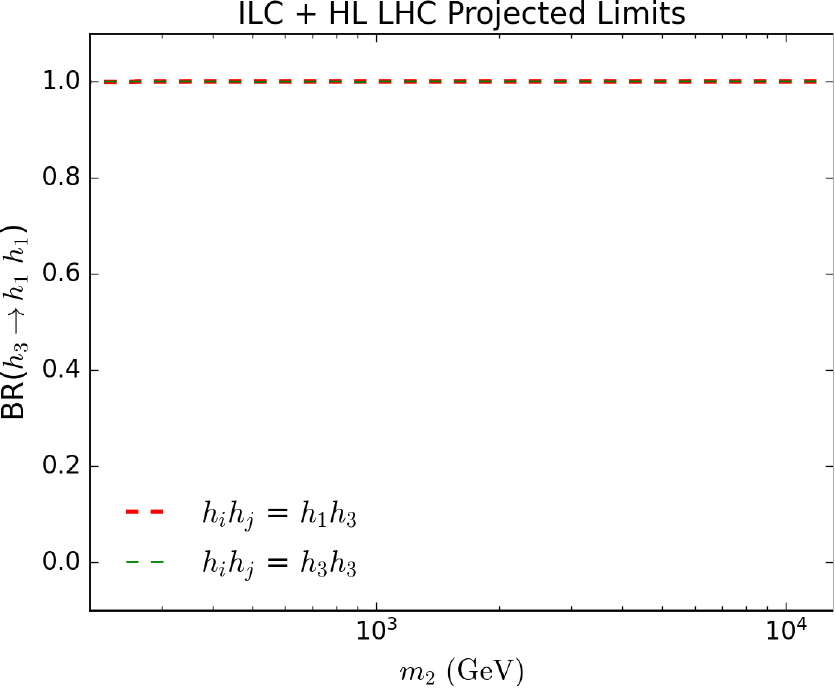}}
\caption{\label{fig:h3h1h1} ${\rm BR}(h_3\rightarrow h_1h_1)$ when the di-scalar resonance rates for (red) $h_2\rightarrow h_1h_3$ and (green) $h_2\rightarrow h_3h_3$ are maximized for (a) scenario S1 using $\chi^2_{\rm Tot}$ fits to scalar searches and Higgs data, (b) scenario S1 using the hard cut method for scalar searches with Higgs precision data fits, (c) scenario S2 with projected HL-LHC limits, (d) scenario S3 with projected FCC-ee+HL-LHC limits, and (e) scenario S4 with projected ILC500+HL-LHC limits.  The mass $m_3=270$~GeV is used and $\theta_2=0.01$.}
\end{center}
\end{figure*}

When kinematically allowed, it is possible to have three and four Higgs signals via $h_3\rightarrow h_1h_1$:  a three Higgs signal from $h_2\rightarrow h_1h_3\rightarrow 3\,h_1$ and four Higgs signals from $h_2\rightarrow h_3h_3\rightarrow 4\,h_1$. We have one benchmark mass $m_3=270$~GeV that allows $h_3\rightarrow h_1h_1$.  In. Fig.~\ref{fig:h3h1h1} we show the branching ratio ${\rm BR}(h_3\rightarrow h_1h_1)$ using the benchmark points that maximize the rates (red) $pp\rightarrow h_2\rightarrow h_1h_3$ and (green) $pp\rightarrow h_2\rightarrow h_3h_3$.  Note that often these two lines overlap and appear as one line.  As can be clearly seen, the branching ratio of $h_3\rightarrow h_1h_1$ is near or at one for the large majority of points when the decay is allowed.  The leading behavior of $\lambda_{113}$ in the small mixing angle $\theta_2$ limit depends on the same combination of potential parameters as $\lambda_{123}$ as seen in Eqs.~(\ref{eq:trilinear113}) and~(\ref{eq:trilinear123}), respectively.  Hence, maximizing $pp\rightarrow h_2\rightarrow h_1h_3$ also maximizes ${\rm BR}(h_3\rightarrow h_1h_1)$.  However, maximizing the rate $pp\rightarrow h_2\rightarrow h_3h_3$ also involves finding points such that $\lambda_{123}=0$.  In this case, the leading behavior of $\lambda_{113}$ is zero and dependence on the mixing angle $\theta_2$ enters.  In Fig.~\ref{fig:h3h1h1}, we set $\theta_2=0.01$. Such a small mixing angle does not impact the our previous conclusions, and the maximum branching ratio of $h_3\rightarrow h_1h_1$ is still near one.  This leads to promising multi-Higgs signals.

The three and four Higgs signals depend on the production rate of $h_1h_3$ and $h_3h_3$ as well as the branching ratio of $h_3\rightarrow h_1h_1$.   In all scenarios, the maximum $pp\rightarrow h_3h_3$ rate is at or above the maximum $pp\rightarrow h_2\rightarrow h_1h_3$ rate.  This leads to the surprising conclusion that the maximum four Higgs production rate is larger or equal to that of the three Higgs production rate. We discuss these signals for each of our scenarios:
\begin{itemize}
\item Scenarios S1 and S2: As seen in Figs.~\ref{fig:S1plots}(a,c) and ~\ref{fig:S2plots}(a), for $m_2\lesssim 600$~GeV, the maximum $h_3h_3$ and $h_1h_3$ rates are similar.  Hence, the maximum triple and quadruple Higgs rates are comparable.  For $h_2$ masses about $600$~GeV, the maximum $h_3h_3$ rate is larger than that of $h_1h_3$.  In this mass range the maximum four Higgs rate is larger than the maximum three Higgs rate.  At $m_2=1$~TeV, the maximum $4h_1$ rate is a factor of $1.6-1.8$ larger than the $3h_1$ rate.  Above 1 TeV, the maximum rate for $pp\rightarrow h_2\rightarrow h_1h_3$ quickly decreases becoming much smaller than the maximum $h_3h_3$ rate.  Hence, the maximum four Higgs rate is much larger than that of the three Higgs, by nearly two orders of magnitude for $m_2\sim\mathcal{O}(10~{\rm TeV})$.
\item Scenarios S3 and S4: For $m_2\lesssim 1$~TeV, the maximum $h_1h_3$ and $h_3h_3$ rates are similar, as shown in Figs.~\ref{fig:S3plots}(a) and~\ref{fig:S4plots}(a).  In this mass range, the maximum three and four Higgs rates will also be similar.  For $h_2$ masses about 1 TeV, the $h_1h_3$ rates decreases quickly becoming much smaller than the maximum $h_3h_3$ rate, as discussed previously.  Hence, similar to scenario S2, the maximum four Higgs rate is much larger than that of the three Higgs in the multi-TeV region.
\end{itemize}
As this discussion makes clear, in the multi-TeV $h_2$ range, this model may be expected to be discovered in the four Higgs channel long before a three Higgs signal may be observed.  Indeed, this model may be discovered in the Higgs pair and quartet production modes.

\section{Conclusion}
\label{sec:conc}
The complex singlet extension to the SM has many spectacular collider signatures in the form of resonant production of multi-scalar final states with a rich phenomenology of decays. In this work, we have explored benchmark scenarios for future collider experiments that maximize production rates of these multi-scalar final states within the complex singlet model. We considered scalar masses of $h_2$ between $250~\gev$ and $12~\tev$, and $h_3$ masses of $130~\gev$, $200~\gev$, and $270~\gev$. 

We found the maximum resonant production rates of $pp\rightarrow h_2\rightarrow h_1h_1/h_1h_3/h_3h_3$ for scenarios including (S1) current constraints on this model and projected constraints from the (S2) HL-LHC, (S3) FCC-ee+HL-LHC, and (S4) ILC500+HL-LHC.  In all scenarios, for $h_2$ masses below 1 TeV, the various maximum rates are comparable with $h_3h_3$ having the largest maximum rate followed by $h_1h_3$ then $h_1h_1$.  As $m_2$ increases into the multi-TeV range, some surprising behavior occurs.  First, the rate for $pp\rightarrow h_2\rightarrow h_1h_3$ and the corresponding branching ratio $h_2\rightarrow h_1h_3$ decrease quickly becoming quite small at 10 TeV.  The maximum rate for $pp\rightarrow h_2\rightarrow h_1h_1$ and $pp\rightarrow h_2\rightarrow h_3h_3$ converge to the same value, with the corresponding branching ratios ${\rm BR}(h_2\rightarrow h_3h_3)\approx 0.5$ and ${\rm BR}(h_2\rightarrow h_1h_1)\approx 0.25$.  We showed that this behavior can be understood analytically through a combination of constraints from requirements of the global minimum, perturbative unitary, narrow width, and Goldstone boson equivalence theorem.  Indeed, analytical formulas were developed to understand the generality of these behaviors.  All benchmark model points corresponding to these maximum rates have been uploaded with the arXiv version of this paper.

The different di-scalar production modes can lead to different collider phenomenology.  While the decays of $h_1h_1$ are always predominantly four $b$-quarks, the $h_1h_3$ and $h_3h_3$ phenomenology depends on the $h_3$ mass.  The $h_3$ mass points $m_3=130$, $200$, and $270$~GeV were chosen to have different collider signatures.  For $m_3=130$ GeV, the dominant decays of $h_3$ are $b\bar{b}$ and $W^\pm W^{\mp,*}$.  Hence, all di-scalar modes preferentially generate multi-$b$ and multi-$W$ signals.  For $m_3=200$~GeV, the dominant decay of $h_3$ is $h_3\rightarrow W^\pm W^\mp$.  Hence, for the asymmetric production $h_1h_3$ the dominate decay is $b\bar{b}W^\pm W^\mp$ and for the symmetric production $h_3h_3\rightarrow 2W^\pm\,2W^\mp$.  That is, we would expect more multi-$W$ signals, particularly in the multi-TeV range where the maximum $h_3h_3$ rate is much larger the maximum $h_1h_3$ rate.

Signals with three and four SM-like Higgs bosons in the final state are possible when $m_3=270$~GeV.  For the parameter points that maximized the $h_1h_3$ or $h_3h_3$ rates we found that ${\rm BR}(h_3\rightarrow h_1h_1)\approx 1$ for the large majority of points. The three Higgs signal originates from $h_1h_3$, whose maximum production rate decreases sharply in the multi-TeV range.  Since the maximum rates for $h_3h_3$ and $h_1h_1$ are similar, we may expect to see both Higgs pair and quartet production when kinematically allowed.  In the multi-TeV range, this model has the surprising feature that it could be discovered in the four Higgs and di-Higgs signals before three Higgs.

\section*{Acknowledgements}
IML would like the Pittsburgh Particle Physics Astrophysics and Cosmology center for
their hospitality during the ongoing work on this paper.  SDL, IML, and MS have been supported in part by the U.S. Department of Energy
Grant No. DE-SC0017988.  SDL,  MS are also supported in part by
the State of Kansas EPSCoR grant program. MS is also supported in part by the United States Department of Energy under Grant Contract DE-SC0012704.  SDL was also supported in part by the National Research
Foundation of Korea (Grant No. NRF-2021R1A2C2009718) and by the University of Kansas General Research Funds. Digital data for the plots and benchmark points have been uploaded with the arXiv version of this paper.

\appendix
\section{Perturbative Unitarity}
\label{app:PertUnit}
The perturbative unitarity matrix in Eq.~(\ref{eq:matrix}) can be block diagonalized by $U\mathcal{M}U^T$ with $U$ given by

\begin{equation}
    U = \begin{pmatrix}
    \frac{1}{2} & 0 & 0 & 0 & 0 & 0 & \frac{1}{2} & -\frac{1}{\sqrt{2}} \\
    \frac{1}{\sqrt{2}} & 0 & 0 & 0 & 0 & 0 & -\frac{1}{\sqrt{2}} & 0 \\
    0 & 1 & 0 & 0 & 0 & 0 & 0 & 0 \\
    0 & 0 & 1 & 0 & 0 & 0 & 0 & 0 \\
    \frac{1}{2} & 0 & 0 & 0 & 0 & 0 & \frac{1}{2} & \frac{1}{\sqrt{2}} \\
    0 & 0 & 0 & 1 & 0 & 0 & 0 & 0 \\
    0 & 0 & 0 & 0 & 1 & 0 & 0 & 0 \\
    0 & 0 & 0 & 0 & 0 & 1 & 0 & 0 \\
    \end{pmatrix}.
\end{equation}
After block diagonalization, $\mathcal{M}$ takes the form
\begin{equation}
    \mathcal{M} = \begin{pmatrix}
    \frac{\lambda}{2} & 0 & 0 & 0 & 0 \\
    0 & \frac{\lambda}{2} & 0 & 0 & 0 \\
    0 & 0 & \frac{\delta_2}{2} + \frac{\Re{\delta_3}}{2} & -\frac{\Im{\delta_3}}{2} & 0 \\
    0 & 0 & -\frac{\Im{\delta_3}}{2} &  \frac{\delta_2}{2} - \frac{\Re{\delta_3}}{2} & 0 \\
    0 & 0 & 0 & 0 & \mathcal{M}_{4}
    \end{pmatrix}
\end{equation}
with
\begin{equation}
    \mathcal{M}_{4} = \begin{pmatrix}
    \frac{3 \lambda}{2} & \frac{\delta_2}{2} + \frac{\Re{\delta_3}}{2} & -\frac{\Im{\delta_3}}{\sqrt{2}} & \frac{\delta_2}{2} - \frac{\Re{\delta_3}}{2} \\ 
    \frac{\delta_2}{2} + \frac{\Re{\delta_3}}{2} & \frac{3 d_2}{4}+\frac{3 \Re{d_1}}{4}+\frac{3 \Re{d_3}}{4} & -\frac{3 \Im{d_1}}{2 \sqrt{2}}-\frac{3 \Im{d_3}}{4 \sqrt{2}} & \frac{d_2}{4} - \frac{3 \Re{d_1}}{4} \\
    -\frac{\Im{\delta_3}}{\sqrt{2}} & -\frac{3 \Im{d_1}}{2 \sqrt{2}}-\frac{3 \Im{d_3}}{4 \sqrt{2}} & \frac{d_2}{2}-\frac{3\Re{d_1}}{2} & \frac{3 \Im{d_1}}{2 \sqrt{2}}-\frac{3 \Im{d_3}}{4 \sqrt{2}} \\
    \frac{\delta_2}{2} - \frac{\Re{\delta_3}}{2} & \frac{d_2}{4} - \frac{3 \Re{d_1}}{4} & \frac{3 \Im{d_1}}{2 \sqrt{2}}-\frac{3 \Im{d_3}}{4 \sqrt{2}} & \frac{3 d_2}{4}+\frac{3 \Re{d_1}}{4}-\frac{3 \Re{d_3}}{4}
    \end{pmatrix}.\label{eq:Mat4}
\end{equation}
A useful rotation to perform on $\mathcal{M}_4$ is
\begin{equation}
    R=
    \begin{pmatrix}
    1 & 0 & 0 & 0 \\
    0 & \frac{1}{2} & \frac{1}{\sqrt{2}} & \frac{1}{2} \\
    0 & -\frac{1}{\sqrt{2}} & 0 & \frac{1}{\sqrt{2}} \\
    0 & \frac{1}{2} & -\frac{1}{\sqrt{2}} & \frac{1}{2}
    \end{pmatrix}
\end{equation}
which yields 
\begin{equation}
    \mathcal{M'}_{4} = \begin{pmatrix}
    \frac{3 \lambda}{2} & \frac{\delta_2}{2} - \frac{\Im{\delta_3}}{2} & -\frac{\Re{\delta_3}}{\sqrt{2}} & \frac{\delta_2}{2} + \frac{\Im{\delta_3}}{2} \\ 
    \frac{\delta_2}{2} - \frac{\Im{\delta_3}}{2} & \frac{3 d_2}{4}-\frac{3 \Re{d_1}}{4}-\frac{3 \Im{d_3}}{4} & \frac{3 \Im{d_1}}{2 \sqrt{2}}-\frac{3 \Re{d_3}}{4 \sqrt{2}} & \frac{d_2}{4} + \frac{3 \Re{d_1}}{4} \\
    -\frac{\Re{\delta_3}}{\sqrt{2}} & \frac{3 \Im{d_1}}{2 \sqrt{2}}-\frac{3 \Re{d_3}}{4 \sqrt{2}} & \frac{d_2}{2}+\frac{3\Re{d_1}}{2} & -\frac{3 \Im{d_1}}{2 \sqrt{2}}-\frac{3 \Re{d_3}}{4 \sqrt{2}} \\
    \frac{\delta_2}{2} + \frac{\Im{\delta_3}}{2} & \frac{d_2}{4} + \frac{3 \Re{d_1}}{4} & -\frac{3 \Im{d_1}}{2 \sqrt{2}}-\frac{3 \Re{d_3}}{4 \sqrt{2}} & \frac{3 d_2}{4}-\frac{3 \Re{d_1}}{4}+\frac{3 \Im{d_3}}{4}
    \end{pmatrix}\label{eq:Mat4p}
\end{equation}
for $\mathcal{M'}_4\equiv R \mathcal{M}_4 R^T$.

\section{Bounds on Cubic and Quartic Potential Parameters}
\label{app:CubicBounds}
We can derive some necessary bounds on combinations of cubics from global minimization.  Global minimization requires that the electroweak minimum has the smallest value for the potential. In other words, 
\begin{equation}
\label{eq:cubicinequality}
V\big((0,v_{\rm EW}/\sqrt{2})^T,0\big) \leq V\big((0,x/\sqrt{2})^T, y + i z\big)
\end{equation} 
for all real field values $x$, $y$, $z$. Looking at certain simple directions in field space can make the analysis tractable and yield bounds on some of the cubics. For example, one can examine the $x=v$, $z=0$ direction, from which the combination of cubics $\textrm{Im}(e_1)-\textrm{Im}(e_2)$ will appear. %
Evaluating the inequality of Eq.~(\ref{eq:cubicinequality}) with two of $x,\,y,\,$ or $z$ fixed and minimizing with respect to the third variable yields necessary, but not sufficient, bounds in the $\theta_2\rightarrow 0$ limit:

\begin{align}
   &x=v,z=0:~&\big| \textrm{Im}(e_1)&-\textrm{Im}(e_2) \big| \leq 3\,m_3\, \big[\textrm{Re}(d_1)-\textrm{Re}(d_3)+d_2\big]^\frac{1}{2} \label{eq:cubicbounds1}\\
&x=v,y=0:   &\big| \textrm{Re}(e_1)&+\textrm{Re}(e_2)\big|
   \leq 3 \big[\big(\textrm{Re}(d_1)+\textrm{Re}(d_3)+d_2\big) \label{eq:cubicbounds2} \\
   &&\times\big(&m_1^2 \sin^2{\theta_1} + m_2^2 \cos^2{\theta_1}\big)\big]^\frac{1}{2} \nonumber\\
&x=v,y=z:&\big| \textrm{Im}(e_1)&+\textrm{Im}(e_2)+\textrm{Re}(e_1)-\textrm{Re}(e_2)   \big| 
   \leq 3 \big[\big(d_2-\textrm{Im}(d_3)-\textrm{Re}(d_1)\big) \label{eq:cubicbounds3} \\
   &&\times\big(&m_1^2 \sin^2{\theta_1} + m_2^2\cos^2{\theta_1} + m_3^2 \big)\big]^\frac{1}{2} \nonumber \\
&x=v,y=-z: &\big| \textrm{Im}(e_1)&+\textrm{Im}(e_2)-\textrm{Re}(e_1)+\textrm{Re}(e_2)    \big| 
   \leq 3 \big[\big(d_2+\textrm{Im}(d_3)-\textrm{Re}(d_1)\big) \label{eq:cubicbounds4} \\ 
   &&\times\big(&m_1^2 \sin^2{\theta_1}^2 + m_2^2\cos^2\theta_2 + m_3^2 \big)\big]^\frac{1}{2}.\nonumber
\end{align}

These four inequalities constrain four independent combinations of the real and imaginary parts of $e_1$ and $e_2$.  The inequalities in Eqs.~(\ref{eq:cubicbounds3},\ref{eq:cubicbounds4}) can be combined, and they become
\begin{eqnarray}
\left|\Im{e_1}+\Im{e_2}\right|&\leq& \frac{3}{2}\left[\left(d_2-\Re{d_1}-\Im{d_3}\right)^{1/2}+\left(d_2-\Re{d_1}+\Im{d_3}\right)^{1/2}\right]\nonumber\\
&&\times\left(m_1^2\sin^2\theta_1+m_2^2\cos^2\theta_1+m_3^2\right)^{1/2}\label{eq:cubicbounds5}\\
\left|\Re{e_1}-\Re{e_2}\right|&\leq& \frac{3}{2}\left[\left(d_2-\Re{d_1}-\Im{d_3}\right)^{1/2}+\left(d_2-\Re{d_1}+\Im{d_3}\right)^{1/2}\right]\nonumber\\
&&\times\left(m_1^2\sin^2\theta_1+m_2^2\cos^2\theta_1+m_3^2\right)^{1/2}\label{eq:cubicbounds6}
\end{eqnarray}

We can remove the dependence on the $d_i$ quartic terms by using bounds on the quartics. These bounds are found from perturbative unitarity and demanding the potential be bounded from below.  Each combination of quartics that appear have the following bounds:
\begin{eqnarray}
0 \leq \textrm{Re}(d_1)-\textrm{Re}(d_3)+d_2 \leq \frac{32 \pi}{3}\label{eq:PertUnit1}\\
0 \leq \textrm{Re}(d_1)+\textrm{Re}(d_3)+d_2 \leq \frac{32 \pi}{3}\label{eq:PertUnit2}\\
0 \leq -\textrm{Im}(d_3)-\textrm{Re}(d_1)+d_2 \leq \frac{32 \pi}{3}\label{eq:PertUnit4}\\
0 \leq \textrm{Im}(d_3)-\textrm{Re}(d_1)+d_2 \leq \frac{32 \pi}{3}.\label{eq:PertUnit3}
\end{eqnarray}
The upper bounds in Eqs.~(\ref{eq:PertUnit1}) and (\ref{eq:PertUnit2}) are a result of placing necessary but not sufficient perturbative unitarity upper bounds on the diagonal terms of Eq.~(\ref{eq:Mat4}).  Similarly, the upper bounds in Eqs.~(\ref{eq:PertUnit4}) and (\ref{eq:PertUnit3}) can be obtained from Eq.~(\ref{eq:Mat4p}).  The lower bound in Eq.~(\ref{eq:PertUnit1}) is found by demanding the potential be bounded from below in the $\phi=S=0$ direction.  Equation~(\ref{eq:PertUnit2}) is from the bounded from below condition in the $\phi=A=0$ direction.  Finally, the lower bounds in Eqs.~(\ref{eq:PertUnit4}) and (\ref{eq:PertUnit3}) are found by demanding the $A=\pm S$ with $\phi=0$ directions are bounded from below.

The perturbative upper bounds here can be combined with the global minimum constraints in Eqs.~(\ref{eq:cubicbounds1},\,\ref{eq:cubicbounds2},\,\ref{eq:cubicbounds5},\,\ref{eq:cubicbounds6}) to obtain a series of upper bounds on the scalar trilinear terms in the potential: 
\begin{eqnarray}
|\Im{e_1}-\Im{e_2}|&\leq& \sqrt{96\pi}m_3\\
|\Im{e_1}+\Im{e_2}|&\leq&\sqrt{96\pi}\left(m_1^2\sin^2\theta_1+m_2^2\cos^2\theta_1+m_3^2\right)^{1/2}\label{eq:Ime1pIme2}\\
|\Re{e_1}-\Re{e_2}|&\leq&\sqrt{96\pi}\left(m_1^2\sin^2\theta_1+m_2^2\cos^2\theta_1+m_3^2\right)^{1/2}\label{eq:Ree1MinRee2}\\
|\Re{e_1}+\Re{e_2}|&\leq& \sqrt{96\pi}\left(m_1^2\sin^2\theta_1+m^2_2\cos^2\theta_1\right)^{1/2}
\end{eqnarray}
These can be further combined to find
\begin{eqnarray}
|\Im{e_{1,2}}|&\leq& \sqrt{96\pi}\left[m_3+\left(m_1^2\sin\theta_1^2+m_2^2\cos^2\theta_1+m_3^2\right)^{1/2}\right]\label{eq:Ime12}\\
|\Re{e_{1,2}}|&\leq&\sqrt{96\pi}\left[\left(m_1^2\sin^2\theta_1+m_2^2\cos^2\theta_1+m_3^2\right)^{1/2}\right.\\
&&\left.+\left(m_1^2\sin^2\theta_1+m_2^2\cos^2\theta_1\right)^{1/2}\right]
\end{eqnarray}

Additional useful lower bounds from vacuum stability can be found by considering different field directions:
\begin{align}
&z=0:&0\leq \lambda\left(d_2+\Re{d_1}+\Re{d_3}\right)-\left(\delta_2+\Re{\delta_3}\right)^2\label{eq:VacA0}\\
&y=0: &0\leq \lambda\left(d_2+\Re{d_1}-\Re{d_3}\right)-\left(\delta_2-\Re{\delta_3}\right)^2\label{eq:VacS0}\\
&y=\pm z:&0\leq \lambda\left(d_2-\Re{d_1}\mp\Im{d_3}\right)-\left(\delta_2\mp\Im{\delta_3}\right)^2\label{eq:VacSA}\\
&x=y=\pm z:&0\leq \lambda+4\left(d_2-\Re{d_1}\mp\Im{d_3}\right)+4\left(\delta_2\mp\Im{\delta_3}\right)
\end{align}
Adding together the $z=(1\pm \sqrt{2})y$ directions yields vacuum stability constraints:
\begin{eqnarray}
&x=0:\quad\quad&0\leq 3\left(d_2+\Im{d_1}\right)-2\left(\Im{d_3}+\Re{d_3}\right)\\
&x\neq0:\quad\quad&0\leq 6\,\lambda\left(d_2+\Im{d_1}\right)-4\,\lambda\left(\Im{d_3}+\Re{d_3}\right)\nonumber\\
&&\quad \quad\quad-\left(\Im{\delta_3}+\Re{\delta_3}-2\delta_2\right)^2
\end{eqnarray}

 The vacuum stability bounds in Eqs.~(\ref{eq:VacA0}-\ref{eq:VacSA}) can be combined with the perturbative unitarity bounds in Eqs.~(\ref{eq:PertUnit1}-\ref{eq:PertUnit3}) to find:
\begin{eqnarray}
\left|\delta_2\pm\Re{\delta_3}\right|\leq\sqrt{\lambda\frac{32\pi}{3}}\label{eq:Delt2pmReDelt3}\\
\left|\delta_2\pm\Im{\delta_3}\right|\leq\sqrt{\lambda\frac{32\pi}{3}}\label{eq:delt2Imdelt3}
\end{eqnarray}
Those can be further combined to find:
\begin{eqnarray}
\left|\delta_2\right|&\leq& \sqrt{\lambda \frac{32\pi}{3}}=\frac{8}{v_{\rm EW}}\sqrt{\frac{\pi}{3}}\left(m_1^2\cos^2\theta_1+m_2^2\sin^2\theta_1\right)^{1/2}\label{eq:Delt2}\\
\left|\Re{\delta_3}\right|&\leq& \sqrt{\lambda \frac{32\pi}{3}}\\
\left|\Im{\delta_3}\right|&\leq& \sqrt{\lambda \frac{32\pi}{3}}\label{eq:Imdelt3}
\end{eqnarray}

Finally, there are additional useful bounds on $\delta_2$ and $\delta_3$ from perturbative unitarity.  The scattering matrix in Eq.~(\ref{eq:matrix}) has a block diagonal submatrix:
\begin{eqnarray}
\begin{pmatrix} \mathcal{M}(h S\rightarrow hS)& \mathcal{M}(h S\rightarrow h A)\\ \mathcal{M}(h A \rightarrow h S)&\mathcal{M}(h A\rightarrow h A)\end{pmatrix}=\frac{1}{2}\begin{pmatrix} \delta_2+\Re{\delta_3} & -\Im{\delta_3}\\ -\Im{\delta_3} & \delta_2-\Re{\delta_3}\end{pmatrix}
\end{eqnarray}
The eigenvalues are
\begin{eqnarray}
\mathcal{M}_\pm = \frac{1}{2}\left(\delta_2\pm |\delta_3|\right).
\end{eqnarray}
Perturbative unitarity then implies
\begin{eqnarray}
16\,\pi\geq \delta_2\pm |\delta_3|\geq -16\,\pi.
\end{eqnarray}

\section{Bounds on Branching Ratios in the Multi-TeV Region}
\label{app:TrilinearBounds}
We now use the results of App.~\ref{app:CubicBounds} to derive upper bounds on scalar trilinear couplings and the branching ratios of $h_2\rightarrow h_ih_j$, where $i,j=1,3$, when $m_2$ is in the multi-TeV region.  These bounds may not be saturated when the full constraints are taken into consideration.  Nevertheless, they can provide a qualitative understanding of the behaviors of the trilinear couplings at high masses.

For the trilinears in Eqs.~(\ref{eq:trilinear112},\ref{eq:trilinear123},\ref{eq:trilinear233}), we are interested in the multi-TeV region for $m_2$.  Our bounds will relevant for future colliders when $\sin\theta_1$ is expected to be small.  Hence, we work in the limit $v_{\rm EW},m_1,m_3\ll m_2$ and $|\sin\theta_1|\ll 1$.

First we maximize ${\rm BR}(h_2\rightarrow h_3h_3)$.  To do this, we maximize the magnitude of $\lambda_{233}$ while minimizing the magnitudes of $\lambda_{123}$ and $\lambda_{112}$.  The combinations of ${\rm Re}(e_1),{\rm Re}(e_2)$ and $\delta_2,{\rm Re}(\delta_3)$ appear with opposite signs in $\lambda_{112}$ and $\lambda_{123}$.  Hence, to obtain larger $|\lambda_{233}|$ while keeping $|\lambda_{112}|$ smaller we choose $\Re{e_1}=-\Re{e_2}$ and $\delta_2 = -\Re{\delta_3}$.  Since $\lambda_{233}$ is independent of ${\rm Im}(e_{1,3})$ and ${\rm Im}(\delta_2)$, $\lambda_{123}$ can be eliminated with by $\Im{e_2}=-3\Im{e_1}$ and $\Im{\delta_3}=0$.  With these parameter choices
\begin{eqnarray}
\lambda_{233}&\rightarrow& \frac{2\sqrt{2}}{3}\cos\theta_1\Re{e_2}+\sin\theta_1\,\delta_2\,v_{\rm EW}\\
\lambda_{123}&\rightarrow& 0\\
\lambda_{112}&\rightarrow& \sin\theta_1\cos^2\theta_1\frac{m_2^2}{v_{\rm EW}}\left(1+2\frac{m_1^2}{m_2^2}\right)
\end{eqnarray}
With $\Re{e_1}=-\Re{e_2}$ and $\delta_2=-\Re{\delta_3}$, the inequalities in Eqs.~(\ref{eq:Ree1MinRee2}) and (\ref{eq:Delt2pmReDelt3}), respectively, give us bounds
\begin{eqnarray}
\mid \Re{e_1}\mid &\leq& \sqrt{24\,\pi}\left(m_1^2\sin^2\theta_1+m_2^2\cos^2\theta_1\right)^{1/2}\\
\mid \delta_2\mid &\leq&\sqrt{\lambda\frac{8\,\pi}{3}}=\frac{4}{v_{\rm EW}}\sqrt{\frac{\pi}{3}}\left(m_1^2\cos^2\theta_1+m_2^2\sin^2\theta_1\right)^{1/2}
\end{eqnarray}
Now, use all the inequalities above to find:
\begin{align}
|\lambda_{233}|&\leq 4\sqrt{\frac{\pi}{3}}\left[2\mid \cos\theta_1\mid\left(m_1^2\sin^2\theta_1+m_2^2\cos^2\theta_1\right)^{1/2}+\mid \sin\theta_1\mid \left(m_1^2\cos^2\theta_1+m_2^2\sin^2\theta_1\right)^{1/2}\right]
\end{align}
Expanding around small mixing angle and small mass with the counting $m_1^2/m_2^2\sim \sin\theta_1$ , we can expand and find to $\mathcal{O}(\sin^{3/2}\theta_1)$
\begin{align}
|\lambda_{233}|&\lesssim 8\sqrt{\frac{\pi}{3}}m_2\left(1+\frac{1}{2}|\sin\theta_1|\frac{m_1}{m_2}\right)\label{eq:lam233Bound}\\
|\lambda_{112}|&\approx |\sin\theta_1|\frac{m_2^2}{v_{\rm EW}}.\label{eq:lam112approx}
\end{align}
Hence, the maximum of $|\lambda_{233}|$ is much larger than $|\lambda_{112}|$ in this limit.  This helps explain why the ${\rm BR}(h_2\rightarrow h_3h_3)$ saturate the generic bound in Eq.~(\ref{eq:BRMaxh3h3}) when $m_2$ is multi-TeV. 

The situation for ${\rm BR}(h_2\rightarrow h_1h_3)$ is considerably different.  First, parameters are chosen to make $\lambda_{233}$ small.  In particular, while $\lambda_{112}$ is suppressed by $\sin\theta_1$, $\lambda_{233}$ has a term that is unsuppressed by small mixing angle.    Since $\lambda_{233}$ depends on different trilinear and quartic potential terms than $\lambda_{123}$, $\lambda_{233}$ can be eliminated with the choice
\begin{eqnarray}
\Re{e_2}=3\,\Re{e_1}~\rm{and}~\delta_2=\Re{\delta_3}.
\end{eqnarray}
This choice leaves $\lambda_{123}$ unchanged.
Now, an upper bound on $|\lambda_{123}|$ can be found with a combination of Eqs.~(\ref{eq:Ime1pIme2},\ref{eq:Ime12},\ref{eq:Imdelt3}) and
\begin{eqnarray}
|3\,{\rm Im}(e_1)+{\rm Im}(e_2)|\leq 2|{\rm Im}(e_1)+{\rm Im}(e_2)|+|{\rm Im}(e_1)-{\rm Im}(e_2)|.
\end{eqnarray}
Then we find:
\begin{align}
|\lambda_{123}|&\leq 4\sqrt{\frac{\pi}{3}}\left\{|\cos\theta_1\sin\theta_1|\left[m_3+2\,\left(m_1^2\sin^2\theta_1+m_2^2\cos^2\theta_1+m_3^2\right)^{1/2}\right]\right.\nonumber\\
&\left.\quad\quad\quad\quad+|\cos^2\theta_1-\sin^2\theta_1|\left(m_1^2\cos^2\theta_1+m_2^2\sin^2\theta_1\right)^{1/2}\right\}.
\end{align}
Now, assuming $m_3^2\sim m_1^2\sim \sin\theta_1 m_2^2$ and $|\sin\theta_1|\ll1$, we expand the inequality:
\begin{align}
|\lambda_{123}|\lesssim 4\sqrt{\frac{\pi}{3}}\left(m_1+2\,m_2|\sin\theta_1|\right)\label{eq:lam123lim}
\end{align}
Again, in this limit, the $h_1-h_1-h_2$ trilinear is given by Eq.~(\ref{eq:lam112approx}).

In the small angle and large $m_2$ regime, we have a bound on the partial width of $h_2\rightarrow h_1h_3$
\begin{align}
\Gamma(h_2\rightarrow h_1h_3)\lesssim \frac{4\,m_2}{3}\left(|\sin\theta_1|+\frac{m_1}{2\,m_2}\right)^2
\end{align}
and approximation for the partial width of $h_2$ into SM final state
\begin{eqnarray}
&\Gamma(h_2\rightarrow {\rm SM})&\approx \Gamma(h_2\rightarrow W^+W^-)+\Gamma(h_2\rightarrow ZZ)+\Gamma(h_2\rightarrow h_1h_1)\\
&&\approx \sin^2\theta_1\frac{m_2^3}{8\,\pi v_{\rm EW}^2},\nonumber
\end{eqnarray}
The upper bound on the $h_2\rightarrow h_1h_3$ partial width grows as a single power of $m_2$ and $h_2\rightarrow {\rm SM}$ grows as a cubic power of $m_2$.  As the mass of $m_2$ increases, the ${\rm BR}(h_2\rightarrow h_1h_3)$ will decrease for a fixed mixing angle.  

To find an upper bound on the rate of $pp\rightarrow h_2\rightarrow h_1h_3$, a maximum $\sin\theta_1$ needs to be found.  Using our narrow width requirement 
\begin{eqnarray}
\Gamma(h_2\rightarrow {\rm SM})+\Gamma(h_2\rightarrow h_1h_3)\leq \frac{m_2}{10}
\end{eqnarray}
we find an upper bound on $\sin\theta_1$:
\begin{eqnarray}
|\sin\theta_1|\lesssim 2\sqrt{\frac{\pi}{5}}\frac{v_{\rm EW}}{m_2}.
\end{eqnarray}
With this upper bound, the upper bound on the $h_2\rightarrow h_1h_3$ rates is
\begin{eqnarray}
{\rm BR}(h_2\rightarrow h_1h_3)\lesssim \frac{32\,\pi}{3}\left(1+\frac{1}{4}\sqrt{\frac{5}{\pi}}\right)^2\frac{v_{\rm EW}^2}{m_2^2}\approx 0.11\left(\frac{m_2}{5~{\rm TeV}}\right)^{-2}.\label{eq:h2h1h3Lim}
\end{eqnarray}
Hence, as $m_2$ increases the upper bound on  ${\rm BR}(h_2\rightarrow h_1h_3)$ while maximizing the production rate very rapidly approaches zero like the inverse square of $m_2$.  Therefore, ${\rm BR}(h_2\rightarrow h_1h_3)$ cannot not saturate the branching ratio of 0.5 in Eq.~(\ref{eq:BRMaxh3h3}).

\section{Experimental Limits}
\label{app:lims}

\subsection{ATLAS Higgs Signal Strengths}
\label{app:ATLASmu}

Since the combinations of Higgs measurements in summer 2022~\cite{ATLAS:2022vkf}, ATLAS has updated their results for $h\rightarrow WW^*$ in the $Wh$ and $Zh$ production modes~\cite{ATLAS-CONF-2022-067} from 36.1 fb$^{-1}$ to 139 fb$^{-1}$.  Here, we describe our method of including the new results.

First, we reproduce the previous ATLAS results. Since we are updating the $WW^*$ final state, we focus on combining the different final state signal strengths.  From Ref.~\cite{ATLAS:2022vkf}
\begin{gather}
\mu^{bb}=0.91^{+0.15}_{-0.14},\quad\mu^{WW}=1.20^{+0.12}_{-0.11},\quad\mu^{\tau\tau}=0.96^{+0.12}_{-0.11},\quad\mu^{ZZ}=1.04^{+0.11}_{-0.10},\nonumber\\
\mu^{\gamma\gamma}=1.09^{+0.10}_{-0.09},\quad\mu^{Z\gamma}=2.10^{+0.98}_{-0.95},\quad\mu^{\mu\mu}=1.21^{+0.62}_{-0.60}.\label{eq:muinit}
\end{gather}
The correlation matrix for these final states is found in the auxiliary data of Ref.~\cite{ATLAS:2022vkf}
\begin{eqnarray}
{\rm Corr}^{\rm fin}_{\rm ATLAS}=\begin{pmatrix}1\quad & 0.02\quad & 0.02\quad& 0.02\quad & 0.04\quad & 0\quad & 0 \\
0.02 & 1 & 0.12 & 0.19 &0.22 & 0.02 & 0.02\\
0.02 & 0.12 & 1 & 0.1 & 0.14 & 0.01 & 0.02 \\
0.02 & 0.19 & 0.1 & 1 & 0.24 & 0.02 & 0.04\\
0.04 & 0.22 & 0.14 & 0.24 & 1 & 0.03 & 0.04\\
0 & 0.02 & 0.01 & 0.02 & 0.03 & 1 & 0.01\\
0 & 0.02 & 0.02 & 0.04 & 0.04 & 0.01 & 1\end{pmatrix}
\end{eqnarray}
in the basis of $(bb,\,WW,\,\tau\tau,\,ZZ,\,\gamma\gamma,\,Z\gamma,\,\mu\mu)$.  From here, a covariance matrix can be defined:
\begin{eqnarray}
{\mathbf \Sigma}^{{\rm fin},ij}=\delta^i\delta^j {\rm Cov}^{{\rm fin},ij}\label{eq:cov}
\end{eqnarray}
where $\delta^i$ are the symmetrized uncertainties.
With this covariance matrix, we have a $\chi^2$:
\begin{eqnarray}
\chi^2 = \left(\bm{\mu}-\bm{\hat{\mu}}\right)({\mathbf \Sigma}^{\rm fin})^{-1}\left(\bm{\mu}-\bm{\hat{\mu}}\right),\label{eq:chi2cov}
\end{eqnarray}
where $\bm{\hat{\mu}}$ is a vector of the signal strength central values in Eq.~(\ref{eq:muinit}) and we assume a universal signal strength.  Minimizing the $\chi^2$ and finding the 1$\sigma$ uncertainty results in
\begin{eqnarray}
\mu=1.05\pm0.06.
\end{eqnarray}
This reproduces the results of Ref.~\cite{ATLAS:2022vkf}, helping to validate this method.

We now move to updating the $h\rightarrow WW^*$ signal strength including the $Wh$ and $Zh$ updates.  From Ref.~\cite{ATLAS:2022vkf} we use
\begin{gather}
\mu^{WW}_{{\rm ggF}+bbh}=1.14\pm0.13,\quad \mu^{WW}_{\rm VBF}=1.13^{+0.19}_{-0.18},\quad\mu^{WW}_{tth+th}=1.64^{+0.65}_{-0.61},
\end{gather}
where ${\rm ggF}$ is gluon fusion production, $bbh$ is production in association with a $b\bar{b}$ pair, ${\rm VBF}$ is vector boson fusion, $tth$ is production in association with a $t\bar{t}$ pair, and $th$ is single top plus Higgs production.  For production in association with a vector boson, we use the results with 139 fb$^{-1}$~\cite{ATLAS-CONF-2022-067}
\begin{gather}
\mu^{WW}_{Wh}=0.45^{+0.32}_{-0.29},\quad \mu^{WW}_{Zh}=1.64^{+0.55}_{-0.47}.
\end{gather}
We use the $h\rightarrow WW^*$ submatrix of the full correlation matrix in the auxiliary date of Ref.~\cite{ATLAS:2022vkf}:
\begin{gather}
{\rm Corr}^{WW}=\begin{pmatrix} 1 & -0.1 & 0.01 & 0.01& 0.02 \\ -0.1 & 1 & 0 & 0 & 0.01\\
0.01& 0 & 1 & -0.05 &-0.01\\
0.01 & 0 & -0.05 & 1 & 0\\
0.02& 0.01 & -0.01 & 0 & 1 \end{pmatrix}.
\end{gather}
in the $({\rm ggF}+bbh,\,{\rm VBF},\,{Wh},\,{Zh},\,tth+th)$ basis.  Using the same method as above, the new $WW$ signal strength is
\begin{eqnarray}
\mu^{WW}=1.10\pm0.09.
\end{eqnarray}
Using this for the fit to the total signal strength, we find
\begin{eqnarray}
\mu_{\rm ATLAS}=1.04\pm0.06.
\end{eqnarray}

\subsection{CMS Higgs Signal Strengths}

CMS has a new measurement of $h\rightarrow b\bar{b}$ in the $t\bar{t}h$ production mode~\cite{CMS-PAS-HIG-19-011} from 35.9 fb$^{-1}$ to 138 fb$^{-1}$ since the reported Higgs combination in summer 2022~\cite{CMS:2022dwd}.  Similar to the ATLAS update in the previous subsection, we provide our method of incorporating this new measurement.  As with ATLAS, we first reproduce the previous CMS results.  From Ref.~\cite{CMS:2022dwd} we have the signal strengths for different final states
\begin{gather}
\mu^{Z\gamma}=2.59^{+1.07}_{-0.96},\quad\mu^{\mu\mu}=1.21^{+0.45}_{-0.42},\quad\mu^{bb}=1.05^{+0.22}_{-0.21},\quad\mu^{\tau\tau}=0.85^{+0.10}_{-0.10},\nonumber\\
\mu^{WW}=0.97^{+0.09}_{-0.09},\quad\mu^{ZZ}=0.97^{+0.12}_{-0.11},\quad\mu^{\gamma\gamma}=1.13^{+0.09}_{-0.09}.
\end{gather}
In the $(Z\gamma,\,\mu\mu,\,bb,\,\tau\tau,\,WW,\,ZZ,\,\gamma\gamma)$ basis, the correlation matrix between the different final state signal strengths is
\begin{eqnarray}
{\rm Corr}^{\rm fin}_{\rm CMS}=\begin{pmatrix}1\quad&0\quad&0\quad&0.02\quad&0.06\quad&\quad 0.02&\quad 0.05\\
0&1&0&0.03&0.05&0.03&0.05\\
0&0&1&0.01&0.01&0&0.01\\
0.02&0.03&0.01&1&0.1&0.1&0.16\\
0.06&0.05&0.01&0.1&1&0.12&0.21\\
0.02&0.03&0&0.1&0.12&1&0.2\\
0.05&0.05&0.01&0.16&0.21&0.2&1\end{pmatrix}
\end{eqnarray}
Using Eqs.~(\ref{eq:cov},\ref{eq:chi2cov}), we find the global signal strength
\begin{eqnarray}
\mu = 0.989\pm0.056.
\end{eqnarray}
This is very close to the reported result of $\mu=1.002\pm0.057$~\cite{CMS:2022dwd}.

We now update the $h\rightarrow b\bar{b}$ signal strength.  From Ref.~\cite{CMS:2022dwd}, the $h\rightarrow b\bar{b}$ signal strengths for different production modes are
\begin{eqnarray}
\mu^{bb}_{ggF}=5.31^{+2.97}_{-2.54},\quad \mu^{bb}_{Wh}=1.26^{+0.42}_{-0.41},\quad \mu^{bb}_{Zh}=0.90^{+0.36}_{-0.34},\quad \mu^{bb}_{tth+th}=0.90^{+0.46}_{-0.44}.
\end{eqnarray}
The correlation submatrix for the $b\bar{b}$ final state in the basis of $({\rm ggF},\,Wh,\,Zh,\,tth+th)$ is
\begin{eqnarray}
{\rm Corr}^{bb}=\begin{pmatrix}1\quad&-0.012\quad&-0.01&0\quad\\
-0.02 & 1 & -0.04 & 0\\
-0.01 & -0.04 & 1 & 0 \\
0 & 0 & 0 &1\end{pmatrix}
\end{eqnarray}
We then replace the $\mu^{bb}_{tth+th}$ signal strength with the new results at 138 fb$^{-1}$~\cite{CMS-PAS-HIG-19-011}
\begin{eqnarray}
\mu^{bb}_{tth}=0.33\pm0.26.
\end{eqnarray}
Using Eqs.~(\ref{eq:cov},\ref{eq:chi2cov}), the new $bb$ signal strength is found to be
\begin{eqnarray}
\mu^{bb}=0.71\pm0.18.
\end{eqnarray}
Combining this with all other final states, we find the new CMS global signal strength
\begin{eqnarray}
\mu_{\rm CMS}=0.96\pm0.06.
\end{eqnarray}

\subsection{Scalar Search Limits}
\label{app:ScalSearch}

\begin{table}[tb]
\resizebox{\textwidth}{!}{
\begin{tabular}{|c|c|c|c|c|}\hline\hline
Experiment & Search & Luminosity & $h_2$ mass range & Ref.\\\hline\hline
ATLAS & $h_2\rightarrow ZZ\rightarrow 4\ell,2\ell2\nu$  in ggF and VBF & 139 fb$^{-1}$  &210-2000 GeV &\cite{ATLAS:2020tlo} \\ \hline

ATLAS & $h_2\rightarrow WW+ZZ\rightarrow2\ell2q,\ell\nu2q$ in ggF and VBF& 139 fb$^{-1}$  & 300-5000 GeV& \cite{ATLAS:2020fry} \\ \hline

\multirow{2}{*}{ATLAS} & \multirow{2}{*}{$h_2\rightarrow WW$  in ggF and VBF} & \multirow{2}{*}{139 fb$^{-1}$}  & 200-3800 GeV for ggF& \multirow{2}{*}{\cite{ATLAS-CONF-2022-066}} \\ 
& & & 200-4000 GeV for VBF & \\\hline

ATLAS & $h_2\rightarrow h_1h_1\rightarrow 4b$ in VBF & 126 fb$^{-1}$ & 260-1000 GeV & \cite{ATLAS:2020jgy}\\\hline

ATLAS & $h_2\rightarrow h_1h_1\rightarrow bbWW$ in ggF& 36 fb$^{-1}$ & 500-3000 GeV&\cite{ATLAS:2018fpd}\\\hline

ATLAS & $h_2\rightarrow h_1h_1\rightarrow \gamma\gamma WW$ in ggF & 36 fb$^{-1}$ & 260-500 GeV&\cite{ATLAS:2018hqk}\\\hline

ATLAS & $h_2\rightarrow h_1h_1 \rightarrow 4W$ in ggF & 36 fb$^{-1}$ & 260-500 GeV & \cite{ATLAS:2018ili}\\\hline

ATLAS & $h_2\rightarrow h_1h_1\rightarrow 4b,bb\tau\tau,bb\gamma\gamma$ in ggF& 126-139 fb$^{-1}$ &250-3000 GeV & \cite{ATLAS-CONF-2021-052}\\\hline

CMS & $h_2\rightarrow ZZ\rightarrow 2\ell2q$ in ggF& 138 fb$^{-1}$ & 450-1800 GeV  & \cite{CMS:2021xor} \\ \hline

CMS & $h_2\rightarrow ZZ\rightarrow 4\ell,2\ell2\nu,2\ell2q$ in ggF and VBF combined  & 36 fb$^{-1}$ &130-3000 GeV & \cite{CMS:2018amk} \\ \hline

CMS & $h_2\rightarrow WW\rightarrow 2\ell2\nu$ in ggF and VBF combined&  138 fb$^{-1}$ & 180-5000 GeV & \cite{CMS-PAS-HIG-20-016} \\ \hline
CMS & $h_2\rightarrow WW\rightarrow 2\ell2\nu,\ell\nu2q$ in ggF and VBF combined& 36 fb$^{-1}$ & 200-3000 GeV & \cite{CMS:2019bnu}\\\hline

\multirow{3}{*}{CMS} & \multirow{3}{*}{$h_2\rightarrow h_1h_1 \rightarrow 4b$ in ggF}&\multirow{3}{*}{36 fb$^{-1}$} &260-1200 GeV & \cite{CMS:2018qmt}\\\cline{4-5}

&  & &750-3000 GeV & \cite{CMS:2017aza}\\\cline{4-5}

 &  &  & 750-3000 GeV& \cite{CMS:2018vjd}\\\hline

CMS & $h_2\rightarrow h_1h_1\rightarrow bb\gamma\gamma$ in ggF & 138 fb$^{-1}$ & 260-1000 GeV&\cite{CMS-PAS-HIG-21-011}\\\hline

CMS & $h_2\rightarrow h_1h_1\rightarrow bbZZ$  in ggF& 36 fb$^{-1}$ &260-1000 GeV & \cite{CMS:2020jeo}\\\hline

CMS &$h_2\rightarrow h_1h_1\rightarrow 4W,4\tau,2W2\tau$ in ggF& 138 fb$^{-1}$ &260-1000 GeV & \cite{CMS:2022kdx}\\\hline

CMS & $h_2\rightarrow h_1h_1\rightarrow bbWW,bb\tau\tau$ in ggF& 138 fb$^{-1}$ & 750-4500 GeV&\cite{CMS:2021roc}\\\hline

CMS & $h_2\rightarrow h_1h_1\rightarrow bbWW$  & 138 fb$^{-1}$ &250-900 GeV & \cite{CMS-PAS-HIG-21-005} \\\hline

\multirow{2}{*}{CMS} & \multirow{2}{*}{$h_2\rightarrow h_1h_1\rightarrow bb\tau\tau$ in ggF}& 138 fb$^{-1}$ & 280-3000 GeV& \cite{CMS:2021yci} \\\cline{3-5}
& & 36 fb$^{-1}$ &250-900 GeV & \cite{CMS:2017hea}\\\hline
\end{tabular}}
\caption{\label{tab:ScalSearch} Scalar searches included in 95\% CL upper limits on $\sin\theta_1$. Here $\ell=e,\mu$.  The first column gives the experiment, the second the search channel, the third the integrated luminosity used in the search, the fourth the heavy resonance mass range, and the fifth the reference. VBF is vector boson fusion production, and ggF gluon fusion.}
\end{table}

In Tab.~\ref{tab:ScalSearch} we provide the heavy scalar searches used to obtain the 95\% CL upper limits on $\sin\theta_1$ as given in Sec.~\ref{sec:exconst}.  In some cases there are overlapping searches from the same experiment.  In those cases, we use whichever result has the strongest observed upper bound on the cross section.

\begin{figure}[tb]
\centering
\subfigure[]{\includegraphics[width=0.45\textwidth,clip]{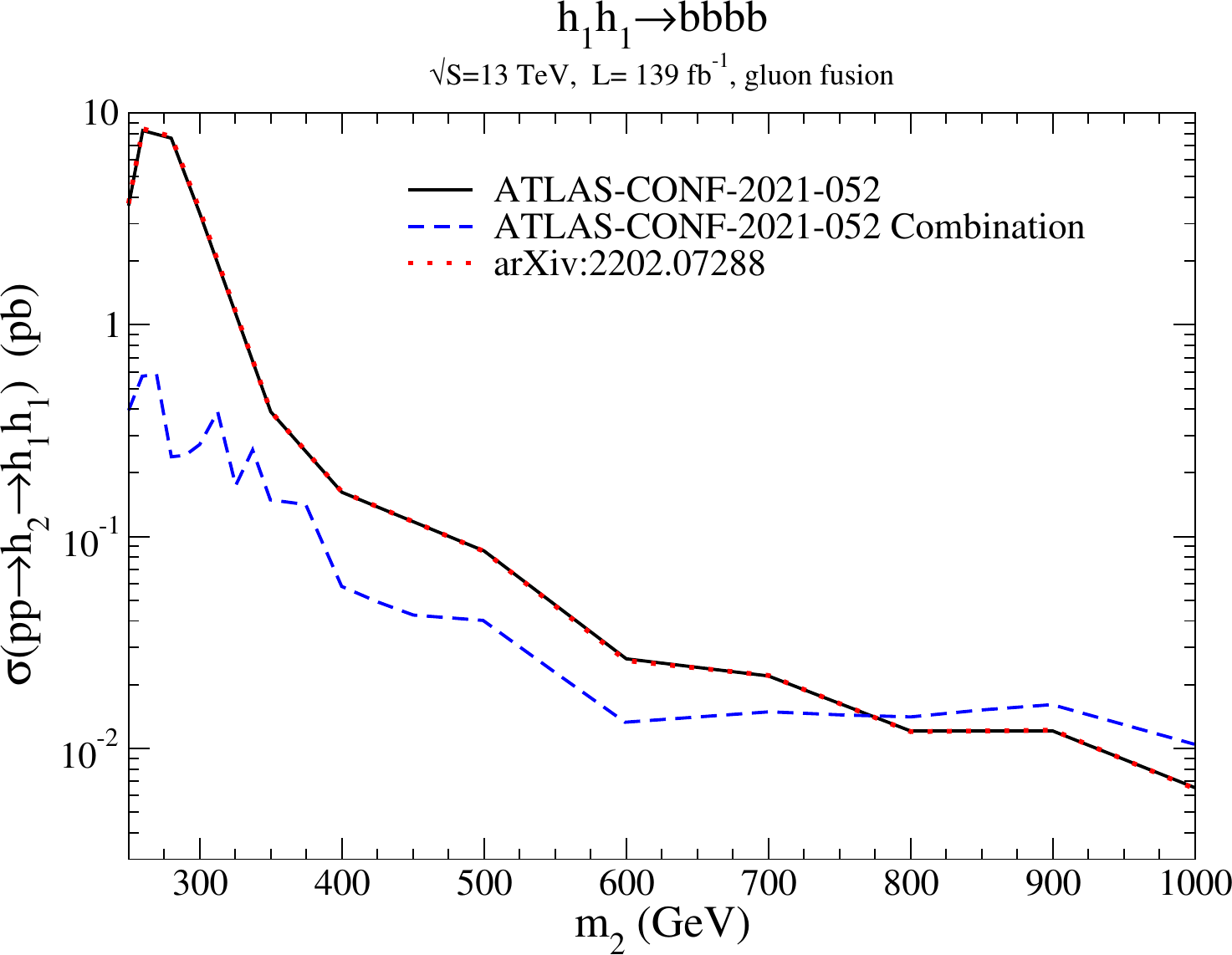}}
\subfigure[]{\includegraphics[width=0.45\textwidth,clip]{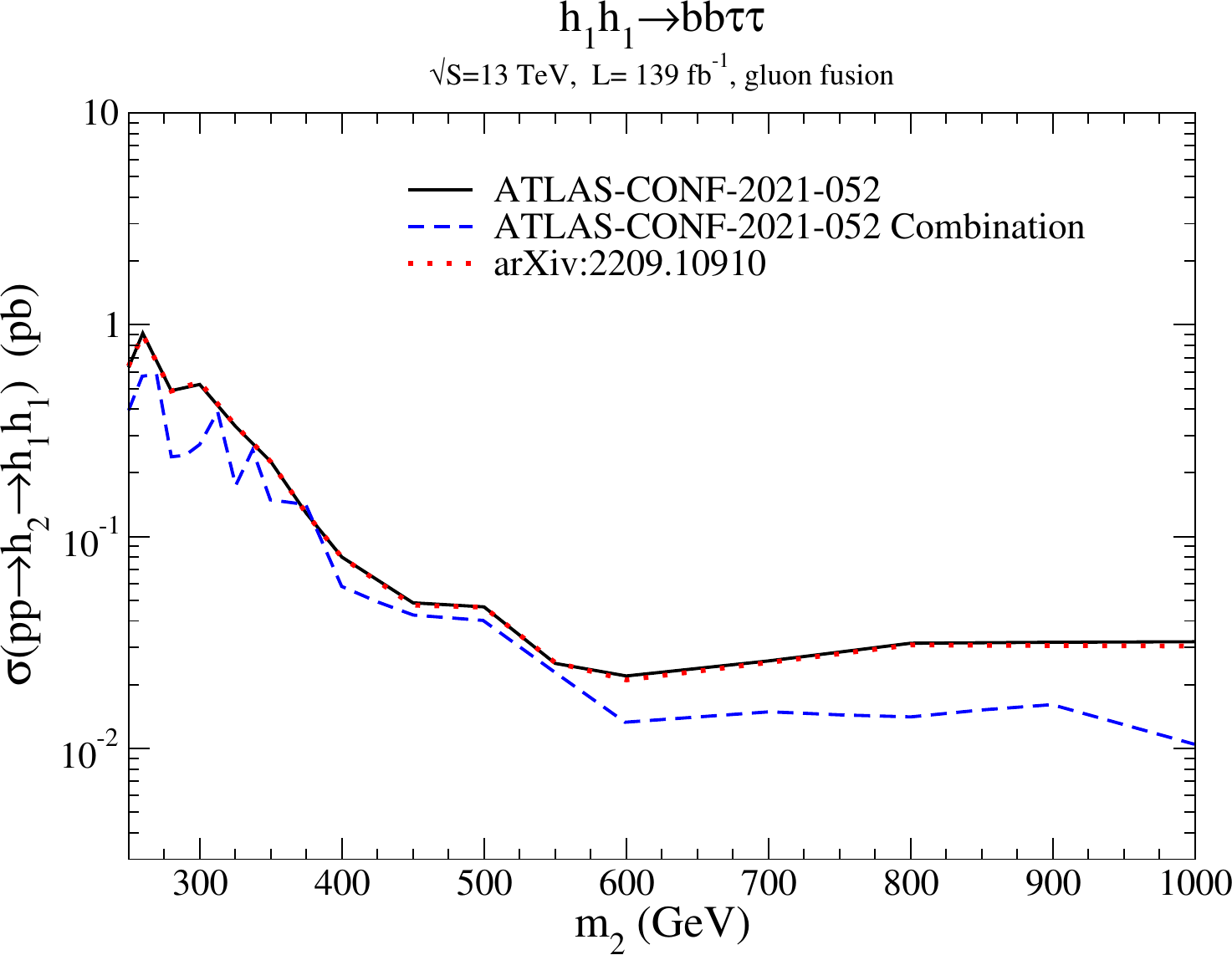}}
\subfigure[]{\includegraphics[width=0.45\textwidth,clip]{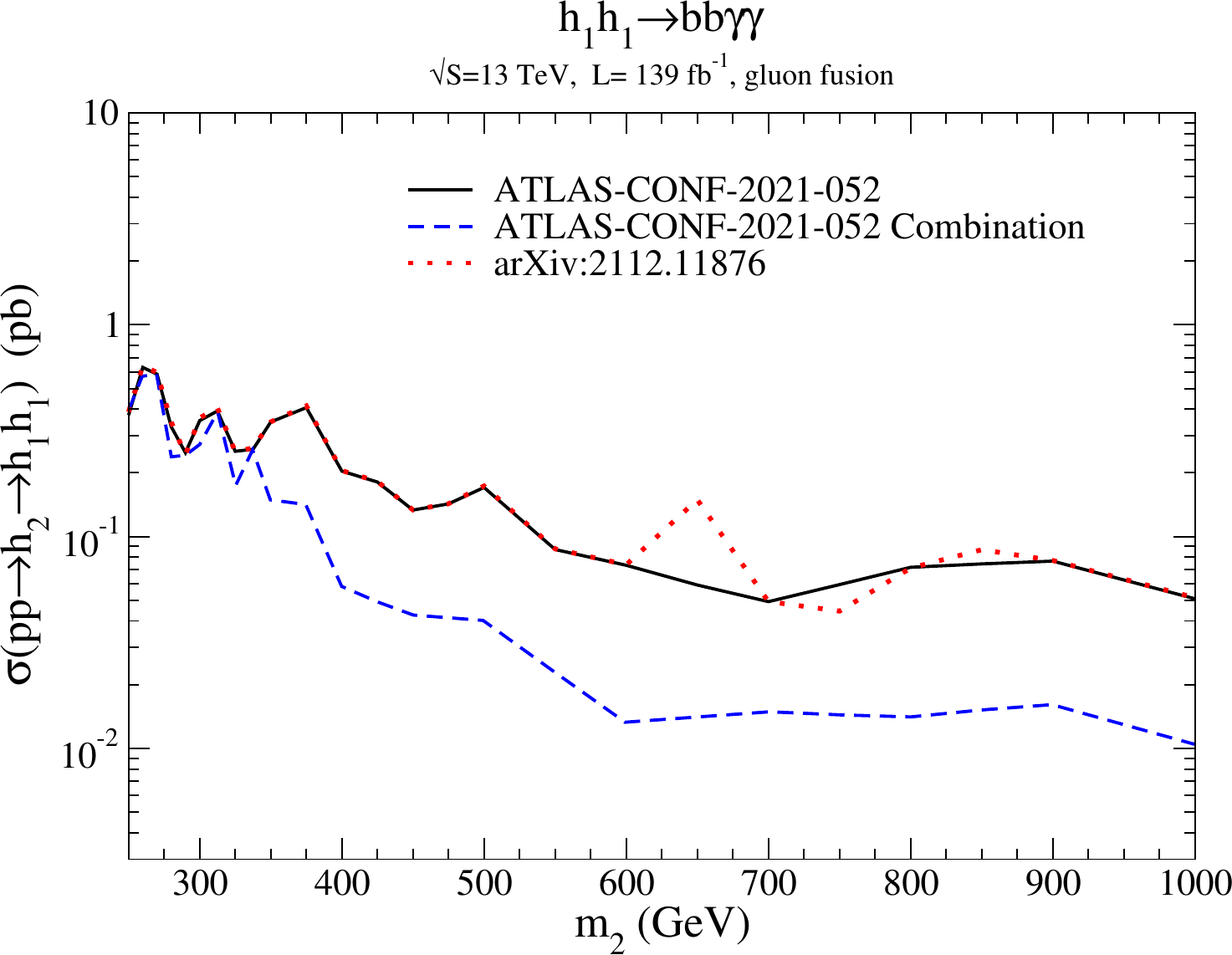}}
\caption{\label{fig:Comparison} Comparison between observed upper bounds on $\sigma(pp\rightarrow h_2\rightarrow h_1h_1)$ reported by ATLAS from the combination~\cite{ATLAS-CONF-2021-052} and individual analysis papers for (a) $4b$~\cite{ATLAS:2022hwc}, (b) $bb\tau\tau$~\cite{ATLAS:2022xzm}, and (c) $bb\gamma\gamma$~\cite{ATLAS:2021ifb} final state.  Solid lines are for the individual channel results from the combination~\cite{ATLAS-CONF-2021-052}, red dotted for the individual analysis papers~\cite{ATLAS:2022hwc,ATLAS:2022xzm,ATLAS:2021ifb}, and the blue dotted for the combination of the channels~\cite{ATLAS-CONF-2021-052}.}
\end{figure}

We use the combined search for $h_2\rightarrow h_1h_1\rightarrow 4b,\,bb\tau\tau,\,b\gamma\gamma$ from ATLAS which appeared in the conference note~\cite{ATLAS-CONF-2021-052}.  There are stand-alone papers for the individual searches of $4b$~\cite{ATLAS:2022hwc}, $bb\tau\tau$~\cite{ATLAS:2022xzm}, and $bb\gamma\gamma$~\cite{ATLAS:2021ifb} with similar luminosity.  Some of these searches appeared after the conference note.  

In Fig.~\ref{fig:Comparison} we compare the individual channels used in the combination~\cite{ATLAS-CONF-2021-052}, those in the individual analysis papers~\cite{ATLAS:2022hwc,ATLAS:2022xzm,ATLAS:2021ifb}, and the combination~\cite{ATLAS-CONF-2021-052}.  First, it should be noted that the $4b$ and $bb\tau\tau$ channels have good agreement between the conference note combination and the individual papers.  

The $bb\gamma\gamma$ analyses have some discrepancies between them at $m_2=650,\,750,$ and $850$~GeV.  These points are not included in the results reported in the combination~\cite{ATLAS-CONF-2021-052}.  The largest discrepancy is at $650$~GeV where there is an upward fluctuation in Ref.~\cite{ATLAS:2021ifb}.  However, these discrepancies occur in mass regions where the $bb\gamma\gamma$ channel is sub-leading to the $4b$ and $2b2\tau$ channel.  Additionally, in the model under consideration here, if there is a significant excess in $bb\gamma\gamma$ there must be an excess in all di-Higgs channels.  This is because the $h_1$ branching ratios are identical to the SM Higgs branching ratios, hence ${\rm BR}(h_2\rightarrow h_1h_1)$ would need to be non-zero, lifting all di-Higgs channels.  With these considerations, we used the combination of these channels in our fits.

\bibliographystyle{utphys}
\bibliography{draft}

\end{document}